\documentclass[12pt,aps,prd, longbibliography, nofootinbib,preprint]{revtex4-1}

\usepackage{braket}
\usepackage{amsmath}
\usepackage{color}
\usepackage{graphicx}
\usepackage{subcaption}
\usepackage{caption}
\usepackage{amssymb}
\usepackage{amsfonts}
\usepackage{tikz}
\definecolor{darkblue}{rgb}{0,0,.5}
\usepackage[colorlinks=true,linkcolor=darkblue,citecolor=darkblue,urlcolor=blue]{hyperref}
\usepackage[format=plain,justification=centerlast,singlelinecheck=true]{caption}
\usepackage{esvect}
\usepackage{bbold}
\usepackage{todonotes}

\usepackage{setspace}

\usepackage[utf8]{inputenc}

\usetikzlibrary{backgrounds,fit,decorations.pathreplacing,calc}

\usepackage[toc,page]{appendix}

\newcommand{\nocontentsline}[3]{}
\newcommand{\tocless}[2]{\bgroup\let\addcontentsline=\nocontentsline#1{#2}\egroup}

\begin{document}

\singlespacing



\title{\Large 
Universality of Black Hole Quantum Computing}

\preprint{LMU-ASC 03/16, MPP-2016-1}

\author{Gia Dvali$^{a, b, c}$}
\author{Cesar Gomez$^{a, d}$}
\author{Dieter L\"ust$^{a, b}$}
\author{Yasser Omar$^{e, f}$}
\author{Benedikt Richter$^{a, e, f}$}

\affiliation{}
\affiliation{$^a$Arnold Sommerfeld Center for Theoretical Physics \\ Department f\"ur Physik, Ludwig-Maximilians-Universit\"at M\"unchen\\Theresienstr. 37, 80333 M\"unchen, Germany}
\affiliation{$^b$Max-Planck-Institut f\"ur Physik \\ F\"ohringer Ring 6, 80805 M\"unchen, Germany}
\affiliation{$^c$Center for Cosmology and Particle Physics \\ Department of Physics, New York University\\ 4 Washington Place, New York, NY 10003, USA}
\affiliation{$^d$Instituto de F\'{i}sica Te\'{o}rica UAM-CSIC \\ Universidad Aut\'{o}noma de Madrid, Cantoblanco, 28049 Madrid, Spain}
\affiliation{$^e$Physics of Information and Quantum Technologies Group\\ Instituto de Telecommunica\c{c}\~oes, Lisboa, Portugal}
\affiliation{$^f$Instituto Superior T\'{e}cnico, Universidade de Lisboa, Portugal\vspace{0mm}}



\collaboration{\small Dated: \today\vspace{10mm}}

\vskip0.5cm

\begin{abstract}

By analyzing the key properties of black holes from the point of view of quantum information, we derive a model-independent picture of black hole quantum computing.  It has been noticed that this picture exhibits striking similarities  with quantum critical condensates, allowing the use of a common language to describe quantum computing in both systems. We analyze such quantum computing by allowing coupling to external modes, under the condition that the external influence must be soft-enough in order  not to offset the basic properties of the system. We derive model-independent bounds on some crucial time-scales, such as the times of gate operation, decoherence, maximal entanglement and total scrambling. We show that for black hole type quantum computers  all these time-scales are of the order of the black hole half-life time.  Furthermore, we construct explicitly a set of Hamiltonians that generates a universal set of quantum gates for the black hole type computer. We find that the gates work at maximal energy efficiency. Furthermore, we establish a fundamental bound on the complexity of quantum circuits encoded on these systems, and characterize the unitary operations that are implementable. It becomes apparent that the computational power is very limited due to the fact that the black hole life-time is of the same order of the gate operation time. As a consequence, it is impossible to retrieve its information, within the life-time of a black hole, by externally coupling to the black hole qubits. However, we show that, in principle, coupling to some of the internal degrees of freedom allows acquiring knowledge about the micro-state. Still, due to the trivial complexity of operations that can be performed, there is no time advantage over the collection of Hawking radiation and subsequent decoding.

\end{abstract}

\maketitle

{\hypersetup{linkcolor=black}
\tableofcontents
}


\section{Introduction and goals}

    In recent years considerable  progress was made in understanding properties of black hole information processing   in terms of universal phenomena characteristic to certain quantum many body systems \cite{Dvali:2011aa, Dvali:2012en}. In the emerging picture  the physics of black hole quantum information processing is understood  in terms of properties of a attractive Bose-Einstein system at the quantum critical point   \cite{Dvali:2012en, Dvali:2015ywa, Flassig:2012re, Dvali:2015wca,  Dvali:2016wca}.  It has been shown that such a critical point is accompanied by the appearance of almost-gapless collective  excitations  that are viable candidates for qubits responsible for the storage and processing of black hole information.  
 
Along this line, recently a more model-independent  and an ``user-friendly" route was taken in  \cite{Dvali:2015wca,  Dvali:2016wca}. This approach can be summarized in the following steps:  
\begin{enumerate}
\item Take only well-established facts about black hole informatics;  
\item Parameterize them in terms  of basic notions of quantum information; 
\item Discover that the black hole quantum informatics parameterized  in this way is in one-to-one correspondence  with properties of  attractive Bose-Einstein condensate at the quantum critical point; 
\item Exploit this correspondence from quantum information perspective (e.g., try to manufacture black hole based quantum computers in the lab).
\end{enumerate} 
  The striking fact that virtually every aspect of black hole informatics finds a counterpart in more ``ordinary" 
  quantum mechanical system has important implications. In particular, it allows to reformulate and analyze 
   black hole type quantum computing in fully-fledged quantum computational language.
   This is the goal of the present work.      
    The power of this approach is that it is absolutely inessential whether 
   one chooses to regard the above-established isomorphy between black holes and the critical condensates as a fundamental 
  connection or only as a remarkable coincidence.   The universality of the language offered by this correspondence is not affected by this choice.   
 
  Since unlike real black holes, their prototypes can be subjected to external manipulation in real laboratory experiments, we shall allow ourselves the analogous theoretical manipulations with the systems of interest, while keeping 
  the bare essentials - which make them isomorphic to black holes - intact.   
   Under such manipulations we mean,  for instance, designing specific logic gates  without affecting 
  properties dictated by quantum criticality, such as the scaling of the energy gap or the strength of the coupling among  different qubits.   This was the approach taken in \cite{Dvali:2015wca, Dvali:2016wca} and the quantum computer designed in this way were referred to as {\it black hole based}  quantum computer.  
  
   The  observer can use the coupling to external modes to read out quantum information stored in the system of  qubits as well as for performing  other logical operations.  It was shown that     
     by coupling critical qubits to external degrees of freedom in this way, one could 
    design simple logic gates, store information in the critical qubits and read out their quantum state. 
    The key point was that thanks to the quantum criticality of the system,  the qubit energy gap $\epsilon$ can be made {\it arbitrarily}  small
    without changing the size of the system.   Correspondingly, the  cost of energy-storage and processing can be made {\it arbitrarily}  cheap.  However, the important result that emerged from this analysis is that the time-scale of any such logical operation performed over an individual qubit - provided 
   this operation respects quantum criticality of the system - takes  a { \it macroscopic }  time given by the inverse 
   of the energy gap, $t \sim \epsilon^{-1}$. 
    When translated to the black hole case, this time-scale is comparable to the black hole half-life time.  
   In other words, the qubits of a quantum critical system are cheap, but take macroscopic time to perform, if the criticality is respected by the process of quantum computation. 
   
    In the present  work we shall generalize this analysis and  investigate the questions of universality, complexity  and efficiency
    of black hole based quantum computers.  
    We shall put ourselves in the position of an observer that can 
  couple the qubits of the critical system (a real black hole or a critical condensate in laboratory) to external degrees of freedom, under the sole condition that the external influence should not offset the quantum criticality of the system.   We shall show that black hole based quantum computing can be made universal, by designing a universal set of gates. 
  
   From the point of view of  efficiency the situation is rather peculiar.  On one hand,  the individual logic gates of a black hole based quantum computer saturate  the efficiency bound  in the sense that they operate within the minimal time compatible with their energy gap, $t\epsilon \sim 1$.      
   Moreover, we can employ a macroscopic number of logic gates working in parallel.  
 So we can perform an unlimited number of  parallel logical operations arbitrarily cheaply.  However,  such a computational process has the following characteristics. 
   The energy used for the quantum computation is a tiny fraction of the total energy of the system. That is, most of the energy serves for maintenance of  criticality.  Moreover, each gate can perform only an order-one operation per half-life time of the system. 
  
     A particular case of a logical operation is the readout  of the black hole quantum state by subjecting its qubits  to an external influence.  Our results,  in agreement with \cite{Dvali:2015wca, Dvali:2016wca},  show that such readout takes the same time as the half-evaporation of the black hole by Hawking radiation.    
      This universal time scale also sets the maximal time for - what we shall refer to as -  {\it local decoherence}. This is the time-scale during which any individual qubit 
      becomes fully entangled with the rest of the qubits.   Thus, this time-scale can also be referred to as the time of maximal entanglement or  time of maximal scrambling.  
      
    To sum it up,  the gate operation time $t_{gate}$,  local decoherence (maximal entanglement) time $t_{decoh}$,   
as well as the system life-time $t_{BH}$, for black hole based quantum computers are set by the same scale, 
\begin{equation}
   t_{gate} \sim t_{decoh}  \sim  t_{BH}  \sim {\epsilon}^{-1} \, .
   \label{ttscales}
 \end{equation}   
       Of course, for the purposes of laboratory quantum computations, we can afford something that the black hole observer cannot: we can externally manipulate the system in and out of quantum criticality. This gives us more potential flexibility for the use of black hole based quantum computations in real labs.

\section{Preliminaries and outline}     
       
There exist certain widely-accepted facts about black hole information properties.  First, there is the existence of the Bekenstein-Hawking entropy \cite{ Bekenstein:1973ur,Hawking:1974sw}, 
 \begin{equation} 
    N \, = \, {R^2 \over L_P^2} \,, 
    \label{entropy} 
  \end{equation}  
  where $R$ is the gravitational radius and $L_P$ is the Planck length. It is well-known that the amount of information corresponding to this entropy, saturates the bound of information-storage capacity of a region of size $R$. Secondly,  according to Page \cite{Page:1993wv}, the time-scale needed for an external observer in order to start the retrieval of information at order one rate, by examining the outgoing Hawking radiation, is  $t_{Page} = N R$. This time-scale is comparable to the half-life time  of a black hole $t_{BH} = NR$.  Another important time-scale is the so-called scrambling time, which was conjectured to be $t_{scrambling} = R\, \ln(N)$ \cite{Sekino:2008he}. Notice, that the parameter $N$ characterizes all the macroscopic properties of the black hole.  In particular,   the black hole mass scales as $M = N {\hbar \over R}$.

 By superimposing the above facts with basic notions of quantum information one can deduce the following model-independent features of quantum qubits that  carry black hole  information.  The  number of qubits scales as $N$.  The energy gap of the qubits scales as $\Delta E_q = {1 \over N} {\hbar \over R}$. The coupling of the qubits  with the other degrees of freedom scales as $\alpha= {1 \over N}$.  In \cite{Dvali:2015wca, Dvali:2016wca}, building on earlier studies
\cite{Dvali:2012en, Dvali:2015ywa, Flassig:2012re, Dvali:2013vxa},  it was demonstrated   that the above macroscopic and microscopic features map on the analogous features of critical Bose-Einstein  systems of attractive bosons at the quantum critical point. The role of the cheap  qubits  is played by nearly-gapless Bogoliubov modes that populate the system at the quantum  critical point.  Thus, an interesting information-theoretical correspondence between critical Bose-Einstein systems and black holes emerges. 

 This striking similarity can be taken as an evidence for the microscopic black hole portrait, according to which black holes  represent the bound-states of $N$-gravitons \cite{ Dvali:2011aa} at the quantum critical point \cite{Dvali:2012en}.   This connection is supported by several studies \cite{Dvali:2012en, Dvali:2015ywa, Flassig:2012re, Dvali:2012uq, Dvali:2012rt,  Dvali:2013vxa, Foit:2015wqa, Dvali:2014ila, Dvali:2015rea, Averin:2016ybl,Dvali:2015wca, Dvali:2016wca},  that reproduce virtually every  aspect of black hole information processing  in terms of critical condensates. In particular, it was shown  that the quantum criticality is crucial for fast scrambling,  due to chaotic behavior and Lyapunov exponent in over-critical regime \cite{Dvali:2013vxa}.  
 
   Even without taking the above microscopic picture literally, the close similarity between the two systems allows to investigate black hole based quantum computing, both theoretically and experimentally, in terms of the critical  Bose-Einstein systems.    This way of parameterizing the computational process allows us to go surprisingly far in understanding black hole information processing. In the present paper we shall adopt this way of thinking. While we shall rely  solely on the above-listed model-independent properties of black hole information qubits, we shall constantly keep in  mind the correspondence  with the critical Bose-Einstein systems.  As a result,  our conclusion shall equally-well apply to both systems.

   Let us first summarize the properties of black hole qubits and logic gates in standard quantum computational terms. The qubit is a two level quantum system, the ground-state and the excited state of which can be denoted by $|0\rangle$ and $|1\rangle$, respectively.   In an appropriate Fock basis, forming a representation of creation annihilation  algebra, $[b, b^{\dagger}] = 1$,  these states can be labeled by the eigenvalues of the occupation number operator $n_b \equiv b^{\dagger}b$.  The first important  characteristic of the qubit is  the energy gap, $ \Delta E_q \, = \, E_1 - E_0 $, between its quantum states $|0\rangle$ and $|1\rangle$.   The scaling of the energy gap already reveals the first peculiarity.  For an ordinary weakly-interacting quantum system of size $R$  (e.g.,  free bosons in box of size $R$) the energy gap between the ground-state and the first excited state would be defined by a minimal uncertainty $\Delta E_{min}$,  which is set by the inverse size of the system.   For example, for the relativistic case,  $\Delta E_{min} \sim {\hbar \over R}$,  whereas in the non-relativistic case, $\Delta E_{min} \sim {\hbar^2 \over mR^2}$, where $m$ is the mass of cold bosons.   
  
  For black holes, as well as for other critical Bose-Einstein condensates, the story is dramatically different: the  gap is suppressed by a macroscopic parameter $N$, 
 \begin{equation}     
     \Delta E \, =\,  \epsilon \,\Delta E_{min},~~ {\rm where}~~ \epsilon \sim {1 \over N} \, .
    \label{gap}
 \end{equation}   
  Given that $N$ is typically a huge number  (e.g.,  for a solar mass black hole, $N \sim 10^{77}$), in  both systems the qubit gap is enormously suppressed as compared to the minimal energy gap $\Delta E_{min}$ that exists in a weakly-coupled system of similar size.  This fact makes the storage of information in Bogoliubov qubits of a critical condensate enormously cheap.

   The second important characteristic of a qubit is the  interaction strength,  $\alpha$, with other qubits and the environment. In the absence of interactions, $\alpha = 0$, the two states, $|0\rangle$ and $|1\rangle$, are the energy eigenstates, and the information carried by the qubit is eternal. In order to store, process and read-out the information, the interaction strength must be non-zero.  The special feature of the black hole qubits as well as the generic Bogoliubov qubits at the quantum critical point is that  the interaction strength is given by $\alpha = {1 \over N}$.  This fact  has far reaching consequences for the  processing of quantum information  by these qubits.   In particular,  any quantum evolution that respects the near-criticality  of the system, gives  a time for an elementary logical operation that saturates the lower bound imposed by the uncertainty  principle. This time scales as $N$. In what follows  we shall work in units $\Delta E_{min} = 1, \hbar =1$.\footnote{These units are much more useful    than $L_P =1$ units, because  $L_P$ is just a cutoff of the theory.}  
  In these units, a typical Hamiltonian, describing a system of two qubits has the following form,   
  \begin{equation}
     H \, := \, \epsilon_b \,  b^\dagger b \,  +  \, \epsilon_c c^\dagger c \, + \, H_{int} \, ,
     \label{Hbc}
 \end{equation} 
 where,  $H_{int}$ is an interaction Hamiltonian.  For example, we can take $ H_{int}\,  = \,  \alpha\, b^\dagger b (c + c^\dagger)$, where  $\alpha$ is the coupling.   This Hamiltonian can describe a two-qubit logic gate, or  an interaction between the qubit and an external field.  Correspondingly, the time evolution can describe a logic gate operation, or a dial-up of a qubit state by  using external radiation. The point is that the maintenance of quantum criticality throughout the unitary evolution demands that $H_{int}$  must not exceed the diagonal terms.   For black holes and critical Bose-Einstein qubits  this bound is saturated \cite{Dvali:2015wca, Dvali:2016wca}, meaning that  during the evolution the  diagonal and off-diagonal terms in the Hamiltonian are of the same order. For example, for black hole qubits  $\epsilon\sim \epsilon_b\sim \epsilon_c \sim \alpha \sim {1\over N}$. Correspondingly, the minimal  time-scale of a unitary evolution, during which the two qubits  can significantly influence each other, scales as 
 \begin{equation}
 t_{gate} \sim {1 \over \epsilon} \, .
 \label{gatetime} 
 \end{equation}   
 This time-scale sets the minimal time-scale of an elementary logical operation that can be performed by the system. For example, this can be a control gate operation during which one of the qubits, acts as a controller of the other one.   Another example is the dial-up of the state vector of the qubit $b$ by means of the interaction with a coherent state of an external radiation mode $c$,  or a read-out of the state of $b$ by scattering at it the external radiation. In both cases the time-scale was shown \cite{Dvali:2015wca, Dvali:2016wca} to scale as (\ref{gatetime}). Notice, for a black hole this time-scale is of the order of the Page time and thus of the order of the black hole half life time.   As it was pointed out, this outcome is fully compatible with the notion that the resolution time of the  black hole quantum hair is of the order of Page's time.  This fact  has very important  consequences for  black hole computational properties.  

In the present paper we wish to characterize the efficiency and universality of a black hole type computation.  The above discussion  shows that individual black hole gates as well as the gates of critical condensate are maximally efficient, in the sense that their speed saturates the bound on evolution-time imposed by the uncertainty principle. Indeed,  equation (\ref{gatetime}) shows that the time-scale of a logical operation is of the order of the inverse qubit energy gap.   Secondly, the memory capacity of the black hole is maximal for a given volume. The black hole type computer can store information in an order-$N$ number of qubits and theoretically perform parallel computations in order-$N$ gates composed by   these qubits.  Given the energy gap of the individual gates $\epsilon \sim {1\over N}$, the total energy invested in storage  of information is $\sim 1$. This is an ${1\over N}$-fraction of the black hole mass, which scales as $M\sim N$.   Thus, a minuscule fraction of the black hole mass  is taken up by the information storage.  However, this should not create an impression that there is a free lunch. Half of the black hole mass is gone during a single logical operation per gate!  So the efficiency of the black hole computation is   limited to performing of order-$N$ logical operations in parallel.  Because of quantum criticality, the  energy gap explored by all the gates collectively is only a $1/N$-fraction of the black hole mass.  However, the entire black hole mass is used as a ``maintenance cost".   In this sense, the black hole is a very short-lived computer (with life-time equal to time of one gate operation), with an extremely cheap information storage and processing, but with enormous maintenance cost.  

This work is organized as follows. In Sec. \ref{bogoinbh}, we give a brief introduction to a toy model of black holes and characterize the collective excitations  that play a crucial role in our studies. In Sec. \ref{uqcinbogoinbh}, we explore whether coupling external modes to these qubits enables one to perform universal quantum computing. Explicitly constructing Hamiltonians that generate a universal set of quantum gates, we find a positive answer to this question. Therefore,  we first show that the Bogoliubov modes of near-critical atomic Bose-Einstein condensates with attractive interactions offer the possibility to perform universal quantum computations. Then in a next step we apply these findings to the collective excitations in black holes and show that black hole type computers are universal quantum computers.  With that in hand, we study in Sec. \ref{seccomp} a fundamental bound on the complexity of quantum circuits in these computers. We find that the complexity of the maximal quantum circuit implementable  is bounded by
\begin{equation}\label{complexityintro}
C_{circuit}^{BH}\sim 1,\hspace{15mm} S_{circuit}^{BH}\sim  N,
\end{equation}
where $C_{circuit}^{BH}$ and $S_{circuit}^{BH}$ denote the circuit depth and size, respectively. The existence of these upper bounds relies on the finite life-time of a black hole and heavily restricts information processing in these systems. Studying the efficiency of computations, we find that each gate is working at the maximal speed, given the amount of energy available for this process. However, a black hole computer provides only a tiny fraction of its energy for this task. Thus, we conclude that most energy is dedicated to the maintenance of criticality. Next, we elaborate on the local decoherence time that we find to be of the same order as the Page-time. Furthermore, considering an external observer that either can directly couple to the black hole qubits or alternatively gather the emitted Hawking radiation, we show that information cannot be retrieved faster by means of externally coupling to the black hole than by the collection of the Hawking radiation.  Finally, we give the conclusions of this work in Sec. \ref{conclusions}.

\section{Bogoliubov modes as qubits}\label{bogoinbh}

The aim of this section is to give an introduction to a simple model that describes 
a Bose-Einstein condensate at the quantum critical point.  As a simple prototype for black hole information-processing, 
this model was first introduced in \cite{Dvali:2012en} and further studied in \cite{Dvali:2015ywa, Flassig:2012re, Panchenko:2015dca, Dvali:2016wca}.\footnote{In the context of cold atomic systems  such models were considered,  e.g., in  \cite{kanamoto2003quantum}, but never from the quantum-computational point of view.  This aspect came into the focus only after the critical systems were identified as the viable black hole prototypes. }   
These studies make it evident that despite of its simplicity,  this model captures the key aspects of black hole information-processing.  Moreover, the explicit parameterization given in \cite{Dvali:2016wca} allows to establish an one-to-one correspondence between the quantum computational properties of the two systems.  
    What we learn from the above analysis is that this correspondence holds due to the universality of the phenomenon of {\it quantum criticality of attractive bosons}, 
 which turns out to be rather insensitive to the concrete nature of bosons (e.g.,  gravitons versus cold atoms with attractive interaction).  
   
Special emphasis is given to the collective excitations (Bogoliubov modes). The occurrence of nearly-gapless, weakly-interacting  Bogoliubov modes near the critical point of a phase transition in Bose-Einstein condensates with attractive interactions is believed to be a generic feature.  In a suitable regime these modes can be described within the Bogoliubov approximation.  In such cases we shall refer to them as Bogoliubov modes. 
 However, the existence of gapless collective excitations near criticality
 extends well beyond the Bogoliubov approximation.
 It is these nearly-gapless excitations that have been identified
 as the key candidates for quantum information storage and processing in  critical systems  \cite{Dvali:2012en, Dvali:2015ywa, Flassig:2012re, Panchenko:2015dca, Dvali:2016wca}.   We shall follow these references.

   The toy model, we consider in the following, is described by the Hamiltonian 
\begin{equation}\label{hamiltonianring}
H_{\Psi}=\int d^d x \Psi^\dagger \frac{-\hbar^2 \Delta}{2m}\Psi-g\hbar\int d^d x \Psi^\dagger \Psi^\dagger\Psi \Psi,
\end{equation}
where $\Psi=\frac{1}{\sqrt{V}}\sum_{\vv{k}} e^{i\frac{\vv{k}\vv{x}}{R}}a_{\vv{k}}$ and $V=(2\pi R)^d$ denotes the volume. The bosonic creation and annihilation operators $a_{\vv{k}}^\dagger$ and $a_{\vv{k}}$ satisfy the usual commutation relations $[a_{\vv{k}},a_{\vv{k'}}]=[a_{\vv{k}}^\dagger,a_{\vv{k'}}^\dagger]=0$ and $[a_{\vv{k}}, a_{\vv{k'}}^\dagger]=\delta_{\vv{k},\vv{k'}}$. In the following, we restrict to $d=1$, i.e., bosons on a ring, where it is known that for some critical value of the coupling $g=g_c$ there is a phase transition \cite{kanamoto2003quantum}. Furthermore, we rescale the Hamiltonian $H_{\Psi}$ by $\frac{2R^2m}{\hbar^2}$ and define $\alpha=\frac{gm}{\pi \hbar R}$. Thus we arrive at the Hamiltonian
\begin{equation}
H=\sum_k k^2 a_k^\dagger a_k-\frac{\alpha}{4}\sum_{k_1,k_2,k_3,k_4} a_{k_1}^\dagger a_{k_2}^\dagger a_{k_3} a_{k_4} \delta(k_1+k_2-k_3-k_4).
\label{Haa}
\end{equation}
To simplify the following considerations we restrict the allowed values of the momentum $k$ to $k=0, \pm 1$. Justification of this procedure for the parameter regime we are interested in was obtained in \cite{Dvali:2015ywa, Flassig:2012re, Panchenko:2015dca}. Further using that the Hamiltonian (\ref{hamiltonianring}) is particle conserving and, thus
\begin{equation}
a_0^\dagger a_0 +a_1^\dagger a_1 +a_{-1}^\dagger a_{-1}=N
\end{equation}
one obtains the following Hamiltonian in the Bogoliubov approximation (i.e., $a_0=a_0^\dagger\approx\langle a_0\rangle \approx\langle a_0^\dagger\rangle\approx \sqrt{N}$)
\begin{equation}
H=\sum_{k=\pm 1} (1+\frac{\alpha N}{ 2}) a_k^\dagger a_k-\frac{\alpha N}{4}\sum_{k=\pm 1} (a_{k}^\dagger a_{-k}^\dagger + a_{k} a_{-k}),
\end{equation}
where we omitted a constant. This quadratic Hamiltonian can be diagonalized using the Bogoliubov transformation
\begin{equation}
a_k=u_k b_k+v^*_k b^\dagger_k
\end{equation}
and one finds the Hamiltonian $H_b$ for the collective excitations 
\begin{equation}\label{lagbogo}
H_b=\epsilon (b^\dagger_{-1}b_{-1}+b^\dagger_{+1}b_{+1}),
\end{equation}
where $N$ is the particle number and $\epsilon=\sqrt{1-\alpha N}$ is the energy gap of the $b-$modes \cite{Dvali:2012en}. Thus, at the critical point ($\alpha N=1$) these modes become massless to lowest order in $\frac{1}{N}$. Including $\frac{1}{N}-$corrections one finds the following effective Hamiltonian $H_b$ describing the Bogoliubov modes  
\begin{equation}\label{lagbogo2}
H_b=\epsilon (b^\dagger_{-1}b_{-1}+b^\dagger_{+1}b_{+1})+\frac{1}{N\epsilon^2}\mathcal{O}(b^4),
\end{equation}
where $\mathcal{O}(b^4)$ denotes the interaction terms between $b-$modes \cite{Dvali:2015wca}.  It is clear that for any given arbitrarily small $\epsilon$ we can always take $N$ sufficiently large in order to make the interaction term
in (\ref{lagbogo2}) irrelevant for states with finite occupation numbers of $b$-modes.   In other words, by taking the double scaling limit,  i.e., $\epsilon\to 0$ and $\epsilon^2 N\to \infty$, we can extend the validity of 
 Bogoliubov Hamiltonian (\ref{lagbogo}) arbitrarily close to quantum critical point. 
 In consequence, the energy gap  $\epsilon$ becomes arbitrarily small and the time evolution of the Bogoliubov modes that is generated by Hamiltonian (\ref{lagbogo}) becomes very slow in this limit. These are the features that make these modes attractive candidates for qubits.  We must note that the existence of nearly-gapless modes in the near-critical regime, $|1 - \alpha N| \ll 1$, is a generic property and holds beyond the validity of  Bogoliubov approximation  \cite{Panchenko:2015dca, Dvali:2015wca, Dvali:2015ywa, Flassig:2012re}.
 Nevertheless, for simplicity we shall restrict ourselves to the Bogoliubov regime.

 Although the appearance of nearly-gapless qubits around the critical point 
 is a universal property of attractive bosons in all dimensions, the number and diversity of gapless qubits is model-dependent and is determined by the type of the attractive interaction.   For example, in the derivatively-coupled toy model considered in \cite{Dvali:2015ywa}, the Hamiltonian $H_{\text{eff}}$ at the critical point up to order $\frac{1}{N}$  is found to be
\begin{equation}\label{hamderivcoupled}
H_{\text{eff}}=\frac{1}{N}\sum_{k\neq 0} |\vv{k}|^2 \left(b_{\vv{k}} ^\dagger b_{\vv{k}} \right)^2.
\end{equation}
Thus, the gap in this case scales like $\epsilon_k \sim\frac{|\vv{k}|^2}{N}$.  Correspondingly, in this case the gapless qubits are labeled by $k$ and  
their number is much larger then in non-derivative case.  

As said above,  Bogoliubov modes described by the Hamiltonian (\ref{hamiltonianring}), (\ref{hamderivcoupled}) exhibit properties that are 
strikingly similar to the ones expected for black hole qubits, briefly discussed in the introduction.  
Their energy gap vanishes for $N\to\infty$ and thus their time evolution governed by (\ref{hamderivcoupled}) becomes very slow. Furthermore, as their interactions are highly suppressed, the creation of entanglement between these modes takes a macroscopically long time that is polynomial in $N$ \cite{Dvali:2015wca}.

After this brief introduction, in the next section, we establish the universality of quantum computing employing near-critical Bogoliubov modes as qubits.

\section{Universal quantum computing with critical Bogoliubov modes}\label{uqcinbogoinbh}

Recently,  the idea of using  critical Bose-Einstein systems as a platform for information processing was put forward in  \cite{Dvali:2015wca}.  It was shown that by coupling Bogoliubov modes ($b-$modes), described by Hamiltonian (\ref{lagbogo}), to some external modes $c$, the state of a $b-$mode, $|0\rangle_b$ or $|1\rangle_b$, can be read out. 
 The crucial point is that if the coupling with the external mode $c$ preserves the near-criticality of the system of Bogoliubov 
 qubits $b$, the minimal possible read-out time is, 
   \begin{equation} 
      t_{read-out} = {1 \over \epsilon} \,. 
   \label{readoutt}
   \end{equation} 
   This simple fact has  important consequences, since, as discussed above,  both for black holes as well as for critical Bose-Einstein condensates 
   the gap $\epsilon$ is suppressed by powers of a macroscopic parameter $N$.  Thus, the time-scale required 
   for resolving the internal quantum state of the system of $b$-qubits is {\it macroscopic}  in $N$.   For example, in case of a black hole the gap scales as $\epsilon = {1 \over N}$, and correspondingly, the resolution 
   time  scales as $t_{read-out} = N$,  which is comparable to the black hole half-life time.  
    Thus, the time-scale required for resolving the quantum state of a black hole qubit by an external measurement is comparable to the time-scale 
     during which the information would leak out by the Hawking radiation, as also discussed in detail in Sec.\ \ref{aspects}.   
    In other words, the quantum hair (i.e., the capacity of read-out of an internal quantum state by an external measurement)  of critical condensates as well as black holes is extremely soft and is suppressed by powers of $N$.  Furthermore, by the same coupling one can prepare $b-$modes in the respective states as well as arrange the 
  system $b$ and $c$ modes in form of logic gate.  Also in this case,  the minimal time required for  
  each operation is given by (\ref{gatetime}),  which agrees with (\ref{readoutt}).

  In the present work we show that universal quantum computing is possible utilizing the systems of critical Bogoluibov modes.   Therefore, we explicitly construct a set of Hamiltonians that generates a universal set of quantum gates using these modes. However, first, to keep this work self-contained, we include a short introduction to universal sets of quantum gates, see \cite{nielsen2010quantum} for details.

\subsection{Universal sets of quantum gates}

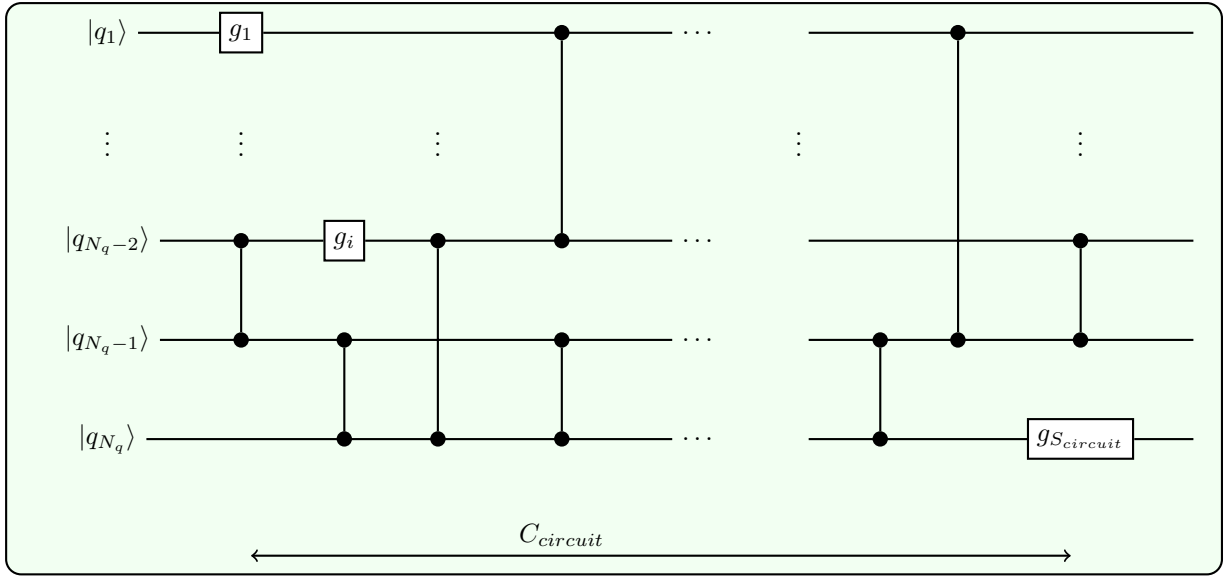
\begin{figure}
\begin{footnotesize} 
\begin{tikzpicture}[thick]
   
    \tikzstyle{operator} = [draw,fill=white,minimum size=1.5em] 
    \tikzstyle{phase} = [draw,fill,shape=circle,minimum size=5pt,inner sep=0pt]
    \tikzstyle{surround} = [fill=green!05,thick, draw=black,rounded corners=2mm]
    \tikzstyle{cnot} = [draw,shape=circle,minimum size=5pt,inner sep=0pt]

  \tikzset{
operator/.style = {draw,fill=white,minimum size=1.5em},
operator2/.style = {draw,fill=white,minimum height=3cm},
phase/.style = {draw,fill,shape=circle,minimum size=5pt,inner sep=0pt},
surround/.style = {fill=green!05,thick,draw=black,rounded corners=2mm},
cross/.style={path picture={ 
\draw[thick,black](path picture bounding box.north) -- (path picture bounding box.south) (path picture bounding box.west) -- (path picture bounding box.east);
}},
crossx/.style={path picture={ 
\draw[thick,black,inner sep=0pt]
(path picture bounding box.south east) -- (path picture bounding box.north west) (path picture bounding box.south west) -- (path picture bounding box.north east);
}},
circlewc/.style={draw,circle,cross,minimum width=0.3 cm},
}

    \matrix[row sep=0.7cm, column sep=0.8cm,] (circuit) {
    \node (q1) {}; &[-0.5cm] 
        \node[] (P11) {$|q_1\rangle$}; &
    \node[operator] (H12) {$g_1$};&
    \node[] (P13) {}; &
    \node[] (P14) {};  &
        \node[phase] (P15) {}; &
    \node[] (P16) {$\dots$};  &
        \node[] (P17) {}; &
    \node[] (P18) {}; &
    \node[phase] (P19) {};  &
        \node[] (P110) {};  &
    \node (P111) {}; &[0.5cm] 
    \coordinate (end1); \\

     \node (q2) {}; &[-0.5cm] 
        \node[] (P21) {$\vdots$}; &
    \node[] (P22) {$\vdots$}; &
    \node[] (P23) {}; &
    \node[] (P24) {$\vdots$};  &
        \node[] (P25) {}; &
    \node[] (P26) {};  &
        \node[] (P27) {$\vdots$}; &
    \node[] (P28) {}; &
    \node[] (P29) {};  &
        \node[] (P210) {$\vdots$};  &
    \node (P211) {}; &[0.5cm] 
    \coordinate (end2); \\
   \node (q3) {}; &
      \node[] (P31) {$|q_{N_q-2}\rangle$}; &
    \node[phase] (P32) {}; &
    \node[operator] (H33) {$g_i$}; &
   \node[phase] (P34) {}; &
    \node[phase] (P35) {}; &
    \node[] (P36) {$\dots$}; &
   \node[] (P37) {}; &
    \node[] (P38) {}; &
       \node[] (P39) {}; &
        \node[phase] (P310) {}; &
          \node (P311) {}; &[-0.5cm] 
    \coordinate (end3); \\

   \node (q4) {}; &
       \node[] (P41) {$|q_{N_q-1}\rangle$}; &
    \node[phase] (P42) {}; &
    \node[phase] (P43) {}; &
   \node[] (P44) {}; &
    \node[phase] (P45) {}; &
    \node[] (P46) {$\dots$}; &
   \node[] (P47) {}; &
    \node[phase] (P48) {}; &
       \node[phase] (P49) {}; &
        \node[phase] (P410) {}; &
          \node (P411) {}; &[-0.5cm] 
    \coordinate (end4); \\

   \node (q5) {}; &
    \node[] (P51) {$|q_{N_q}\rangle$}; &
        \node[] (P52) {}; &
    \node[phase] (P53) {}; &
   \node[phase] (P54) {}; &
    \node[phase] (P55) {}; &
    \node[] (P56) {$\dots$}; &
   \node[] (P57) {}; &
    \node[phase] (P58) {}; &
       \node[] (P59) {}; &
        \node[operator] (H510) {$g_{S_{circuit}}$}; &
          \node (P511) {}; &[-0.5cm] 
    \coordinate (end5); \\

   \node (q6) {}; &
    \node[] (P61) {}; &
       \node[] (P62) {}; &
          \node[] (P63) {}; &
             \node[] (P64) {}; &
                \node[above] (P65) {$C_{circuit}$};
                   \node[] (P66) {}; &
                      \node[] (P67) {}; &
                         \node[] (P68) {}; &
                            \node[] (P69) {}; &
                             \node[] (P610) {}; &
                            
          \node (P611) {}; 
          \node (P612) {}; &[-0.5cm] 
    \coordinate (end6); \\
   };

   \draw[thick]  (P11) -- (H12)  (H12) -- (P16) (P31) -- (H33) (H33) -- (P36) (P41) -- (P46) (P51) -- (P56)  (P17) -- (P111)  (P37) -- (P311) (P47) -- (P411)  (P57) -- (H510)  (H510) -- (P511)  (P32) -- (P42)  (P43) -- (P53)  (P54) -- (P34)  (P55) -- (P45)  (P35) -- (P15)   (P48) -- (P58)   (P49) -- (P19)   (P310) -- (P410)  ;
   
   \draw[thick,<->] (P62) -- (P611)  ;

    \begin{pgfonlayer}{background}

       \node[surround] (background) [fit = (H12) (q6) (P311)] {};
               
  \end{pgfonlayer}
    \end{tikzpicture}
    \end{footnotesize} 
\caption{Schematic picture of the circuit size $S_{circuit}$ and the circuit depth $C_{circuit}$. There are $C_{circuit}$ time steps and in total  $S_{circuit}$ quantum gates, where at most $N_q$ gates can be applied in parallel. Here, the $g_i$ denote some gate operation and the qubits are labeled as $q_j$. Any unitary operation $U$ on $N_q$ qubits can be approximated to arbitrary precision by such a quantum circuit. However, the number of gates required  might be an exponential function of $N_q$.} \label{figcircuit}
\end{figure}

A quantum (logic) gate $g^{(i)}$ is a unitary $2^i \times 2^i-$matrix $U_g^{(i)}$ acting on $i$ qubits in state $|i\rangle$, as 
\begin{equation}
|i\rangle\to |i'\rangle =U_g^{(i)} |i\rangle
\end{equation} 
and therefore is reversible. A set  of quantum gates $G=\{g_1^{(i_1)}, g_2^{(i_2)},\dots , g_n^{(i_n)}\}$ is called universal if for all $\epsilon>0$ and all unitary operations $U$ on a number $N_q$ of qubits there exists a sequence of gates (a circuit) such that
\begin{equation}
\max_{||\,|v\rangle\,||=1} ||\left(U-U_{g_{j_1}}U_{g_{j_2}}\dots U_{g_{j_l}}\right)|v\rangle||\leq \epsilon,
\end{equation}
where we dropped the label signalizing the number of qubits a gate is acting on and $j_i\in \{1,2,\dots, n\}$. One universal set of two-qubit universal gates is given by the Hadamard gate $H$, the $R(\frac{\pi}{4})$ gate and the controlled $NOT$ ($CNOT$) gate, cf. Appendix \ref{quantumgates}. Note that the Hadamard gate $H$ and the $R(\frac{\pi}{4})$ gate are one-qubit universal and that any one-qubit universal set of gates can be made universal for any number of qubits by including almost any two-qubit gate \cite{Deutsch:1995dw, Lloyd:1995zz} (the constraint is that it has to be an entangling gate). Therefore, any unitary operation can be approximated by these gates. 

For concrete realizations an important question is how many gates are needed to approximate a given unitary matrix, i.e., the complexity of the circuit. A generic unitary operator on $n$ qubits has $2^n$ parameters and therefore cannot be efficiently approximated by a universal set of gates. \footnote{We say a circuit efficiently approximates a unitary operator on $n$ qubits if the circuit consists of a  polynomial (in $n$) number of gates.} The Solovay-Kitaev theorem states that an arbitrary one-qubit gate can be approximated with accuracy $\epsilon$ by $\mathcal{O}(\log^2(\frac{1}{\epsilon}))$ gates. However, for a generic unitary the number of gates required to approximate it is of the order $\mathcal{O}(2^n\log^2(\frac{1}{\epsilon})/\log(n))$. Therefore, there are some unitaries that cannot be approximated efficiently \cite{nielsen2010quantum}.

To formalize the circuit complexity, we introduce the circuit size $S_{circuit}$ as the total number of gates  and the circuit depth $C_{circuit}$ as the number of time steps, see Fig.\ \ref{figcircuit}. These quantities depend on the size of the input (number of qubits). We call unitary operations that can be approximated by circuits that have polynomial size and depth efficiently approximable. That is, if
\begin{equation}
C_{circuit}=\text{poly}_2(N_q),\hspace{10mm} S_{circuit}=\text{poly}_1(N_q)
\end{equation}
where $\text{poly}_i(N_q)$ ($i=1,2$) are some polynomials of the number of input qubits, $N_q$, we call the circuit efficient. In most cases, one is interested in efficient circuits not because of fundamental space or time limitations, but because of practical limitations. However, studying quantum information in black holes one runs unavoidably into limitations of the time available for a computation, given by the evaporation time. The maximal circuit depth and therefore also the maximal circuit size is, as we see, fundamentally limited in the case of black holes. Next, we construct the Hamiltonians that dynamically generate a universal set of quantum gates for Bogoliubov modes.

\subsection{Hamiltonians generating quantum gates}\label{hamforgates}

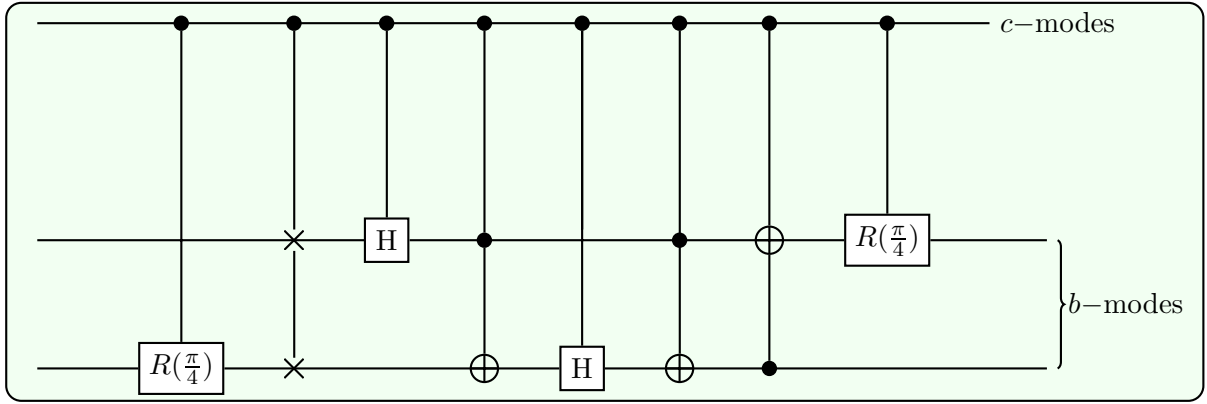
\begin{figure}
\center
\begin{small}
\begin{tikzpicture}[thick]

    \tikzstyle{operator} = [draw,fill=white,minimum size=1.5em] 
    \tikzstyle{phase} = [draw,fill,shape=circle,minimum size=5pt,inner sep=0pt]
    \tikzstyle{surround} = [fill=green!05,thick, draw=black,rounded corners=2mm]
    \tikzstyle{cnot} = [draw,shape=circle,minimum size=5pt,inner sep=0pt]

  \tikzset{
operator/.style = {draw,fill=white,minimum size=1.5em},
operator2/.style = {draw,fill=white,minimum height=3cm},
phase/.style = {draw,fill,shape=circle,minimum size=5pt,inner sep=0pt},
surround/.style = {fill=green!05,thick,draw=black,rounded corners=2mm},
cross/.style={path picture={ 
\draw[thick,black](path picture bounding box.north) -- (path picture bounding box.south) (path picture bounding box.west) -- (path picture bounding box.east);
}},
crossx/.style={path picture={ 
\draw[thick,black,inner sep=0pt]
(path picture bounding box.south east) -- (path picture bounding box.north west) (path picture bounding box.south west) -- (path picture bounding box.north east);
}},
circlewc/.style={draw,circle,cross,minimum width=0.3 cm},
}

    \matrix[row sep=1cm, column sep=0.8cm,] (circuit) {

    \node (q1) {}; &[-0.5cm] 
    \node[] (P11) {}; &
    \node[phase] (P12) {}; &
    \node[phase] (P13) {};  &
        \node[phase] (P14) {}; &
    \node[phase] (P15) {}; &
    \node[phase] (P16) {};  &
        \node[phase] (P17) {}; &
    \node[phase] (P18) {}; &
    \node[phase] (P19) {};  &
        \node[] (P110) {$c-$modes};  &
    \node (P111) {}; &[0.5cm] 
    \coordinate (end1); \\

    \node (q2) {}; &[-0.5cm] 
    \node[] (P21) {}; &
    \node[] (P22) {}; &
    \node[] (P23) {}; &
        \node[] (P24) {}; &
    \node[] (P25) {}; &
    \node[] (P26) {}; &
        \node[] (P27) {}; &
    \node[] (P28) {}; &
    \node[] (P29) {}; &
        \node[] (P210) {}; &
     \node (P211) {}; &[0.5cm] 
    \coordinate (end2);\\

   \node (q3) {}; &
    \node[] (P31) {}; &
    \node[] (P32) {}; &
   \node[crossx] (P33) {}; &
    \node[operator] (H34) {H}; &
        \node[phase] (P35) {}; &
    \node[] (P36) {}; &
   \node[phase] (P37) {}; &
    \node[circlewc] (P38) {}; &
       \node[operator] (H39) {$R(\frac{\pi}{4})$}; &
        \node[] (P310) {}; &
          \node (P311) {}; &[-0.5cm] 
    \coordinate (end3); \\

   \node (q4) {}; &
    \node[] (P41) {}; &
    \node[operator] (H42) {$R(\frac{\pi}{4})$}; &
   \node[crossx] (P43) {}; &
    \node[] (P44) {}; &
        \node[circlewc] (P45) {}; &
    \node[operator] (H46) {H}; &
   \node[circlewc] (P47) {}; &
    \node[phase] (P48) {}; &
       \node[] (P49) {}; &
        \node[] (P410) {}; &
          \node (P411) {}; &[-0.5cm] 
    \coordinate (end4); \\
   };

   \draw[thick] (q1) -- (P110)   (q3) -- (H34) (H34) -- (H39)  (H39)--(P310)   (q4) -- (H42)  (H42) -- (H46)  (H46) -- (P410) (P14) -- (H34) (P15) -- (P35) (P16) -- (P36) (P18) -- (P38) (P19) -- (H39) (P12) -- (H42) (P17) -- (P47) (P16) -- (H46) (P13) -- (P33)  (P35) -- (P45)  (P33) -- (P43)   (P38) -- (P48);
   
    \draw[decorate,decoration={brace},thick]
       ($(P310)$)
       to node[midway,right] (bracket) {$b-$modes}
        ($(P410)$);
    \begin{pgfonlayer}{background}
   
       \node[surround] (background) [fit = (q1) (H46) (bracket)] {};
               
   \end{pgfonlayer}
    \end{tikzpicture}
\caption{Schematic picture of the implementation of a unitary transformation on the Bogoliubov modes $b$ controlled by external modes $c$. The circuit is such that all gates acting on $b-$modes are controlled by $c-$modes. The first four gates from the left are  (controlled) $R(\frac{\pi}{4})$ gate, (controlled) $SWAP$ gate, (controlled) Hadamard gate and a Toffoli gate ((controlled) $CNOT$ gate); cf. Appendix \ref{quantumgates}. In that way any circuit can be implemented on the $b-$modes.} \label{figgates}
\end{small}
\end{figure}
The key idea  to achieve the implementation of unitary transformations on $b-$modes of a Bose-Einstein condensate, is to make all gates on $b-$modes controlled by some external $c-$modes. The goal is to find a universal set of quantum gates that  either only acts on $b-$modes but is controlled by external $c-$modes or acts on modes $b$ and $c$; see Fig.\ \ref{figgates}. In order to describe the action of quantum gates on Bogoliubov modes in a Bose-Einstein condensate, we first work in the Bogoliubov regime to establish the Hamiltonians that generate quantum gates. The gates will be implemented by coupling the system to an external system. In the following, we refer to the modes of the system as $b-$modes and to the modes of the external system as $c-$modes. Introducing interactions between these systems, the time evolution is no longer described by Hamiltonian (\ref{lagbogo}), but instead by the modified Hamiltonian
\begin{equation}\label{lagbogomod}
H=\sum_i \epsilon b^\dagger_{i}b_{i}+\chi(t) H_{gate},
\end{equation}
where $H_{gate}$ is the Hamiltonian describing the interactions introduced to generate the gate operation and we dropped the quartic interaction terms that can be made negligible in the double scaling limit. Furthermore, we relabeled the Bogoliubov modes, as $b_{+1}\equiv b_1$ and $b_{-1}\equiv b_2$. The coupling $\chi(t) $ describes a smooth switching function that we will not specify further. To construct our set of gates, we start with the simplest one, the (controlled) phase gate. The construction of the remaining gate Hamiltonians are given in Appendix \ref{constrhamiltonians}. As the phase gate is a single-qubit gate regarding the $b-$modes, we do not specify the label $i=1,2$ and instead call the mode just $b$. A phase gate on mode $b$ is realized by the time-evolution with respect to $H_{\frac{\pi}{4}}$ given by 
\begin{equation} \label{ansatzphase}
H_{\frac{\pi}{4}}=\alpha c^\dagger c b^\dagger b.
\end{equation}
This can be seen as follows. Calculating the action of $H_{\frac{\pi}{4}}$ on the basis states $|0_c 0_b\rangle$, $|0_c 1_b\rangle$, $|1_c 0_b\rangle$ and  $|1_c 1_b\rangle$
\begin{align}
H_{\frac{\pi}{4}} |0_c 0_b\rangle=& 0,  \hspace{20mm}  H_{\frac{\pi}{4}} |1_c 0_b\rangle= 0,\nonumber\\
H_{\frac{\pi}{4}} |0_c 1_b\rangle=& 0,  \hspace{20mm}  H_{\frac{\pi}{4}} |1_c 1_b\rangle= \alpha |1_c 1_b\rangle,
\end{align}
we see that the only state, exhibiting a non-trivial time evolution with respect to $H_{\frac{\pi}{4}}$, is $|1_c 1_b\rangle$. However, including the free evolution of the $b-$modes, we see that after an time $t$ the basis evolved to
\begin{align}
 |0_c 0_b\rangle(t)=& |0_c 0_b\rangle,   \hspace{20mm} |1_c 0_b\rangle(t)= |1_c 0_b\rangle,\nonumber\\
 |0_c 1_b\rangle(t)=& e^{i\epsilon t}|0_c 1_b\rangle,   \hspace{15mm}  |1_c 1_b\rangle(t)= e^{i(\alpha+\epsilon) t} |1_c 1_b\rangle .
\end{align}
Therefore, the action of Hamiltonian (\ref{lagbogomod}) generates a (controlled) phase gate  on the basis states $|0_c 0_b\rangle$, $|0_c 1_b\rangle$, $|1_c 0_b\rangle$ and  $|1_c 1_b\rangle$ in a time $t=\frac{2\pi}{\epsilon}$, where the state $|1_c 1_b\rangle$ picks up the phase $\phi=\frac{2\pi\alpha}{\epsilon}$. Thus, for the choice $\alpha= \frac{\epsilon}{8}$, the $R(\frac{\pi}{4})-$gate is implemented. It is the interaction with external $c-$modes that effectively implements the gate. Similarly, although less straight forward, works the construction of the gate Hamiltonians that complement (\ref{ansatzphase}) to a set of Hamiltonians generating a universal set of quantum gates on $b-$modes, given in Appendix \ref{constrhamiltonians}.

The set of Hamiltonians $\{H_{gate}\}$ that generates controlled operations on $b-$modes with $c-$modes being the control qubits that is universal on the subset of $b-$modes is generated by Hamiltonians (\ref{ansatzphase}), (\ref{ansatzhada}), (\ref{ansatzcnotdv}) and (\ref{ansatzcnot}), given by
\begin{itemize}
\item[1.] Phase gate ($R(\frac{\pi}{4})$):
\begin{equation}
H_{\frac{\pi}{4}}=\frac{\epsilon}{8} c^\dagger c b^\dagger b.
\end{equation}
\item[2.] Hadamard gate ($H$): 
\begin{equation}
H_H=\frac{\epsilon}{4\sqrt{2}} c^\dagger c \left((3\sqrt{2}+1)-2(2\sqrt{2}+1)  b^\dagger b+(b+b^\dagger)\right).
\end{equation}
\item[3.] Controlled $NOT$ gate ($CNOT$): 

The $CNOT$ gate can be obtained from a Toffoli gate $TOFF$ that acts on two $b$-modes controlled by a $c$-mode. The Toffoli gate can be constructed from the previous ones plus $CNOT$ gates that act on one $b$-mode and one $c$-mode  (see Fig.\ \ref{figtoffoli} in Appendix \ref{constrhamiltonians}): 
\begin{align}
H_{CNOT{bc}}=&\frac{\epsilon}{2} c^\dagger c \left(\frac{3}{2}- 2 b^\dagger b+(b+b^\dagger)\right),\\
H_{CNOT_{cb}}=&\epsilon b^\dagger b (c^\dagger +c).
\end{align}

\end{itemize}
All these Hamiltonians generate the respective gate operations during an operation time,
\begin{equation}\label{tgatebec}
t_{gate}^{BEC}\sim \epsilon^{-1}\,.
\end{equation}

\subsection{Universal quantum computing in atomic Bose-Einstein condensates}\label{uqcatomic}

In the previous section, we gave a set of Hamiltonians that creates a set of quantum gates for Bogoliubov modes in an atomic Bose-Einstein condensates with attractive interactions. The set of gates is universal and therefore we can conclude at this point that ``black hole type quantum computing'', as it was coined in \cite{Dvali:2015wca}, is universal. This is, to our knowledge, the first time that the universality for quantum information of atomic Bose-Einstein condensates with attractive interactions utilizing Bogoliubov modes as qubits is discussed. In \cite{Dvali:2015wca} and \cite{Dvali:2016wca} methods are described how the initial state of $b-$modes can be prepared using coherent states. Using this prescription, given an unknown state $|u\rangle$ of a $b-$mode, an eigenstate $|m_0\rangle$ of the number operator $b^\dagger b$ with eigenvalue $m_0$ can be prepared. Subsequently, a circuit built from gates found in Sec.\ \ref{hamforgates} can be applied and finally the readout can be performed using, for example, a $SWAP$ gate between $b-$ and $c-$modes. Alternative methods are described in \cite{Dvali:2015wca, Dvali:2016wca}. Therefore, now the whole tool box of setting up the initial state, performing (universal) computations and the final read out process is available.

Finally, to give a simple example of a quantum circuit implementing a unitary operation on two $b-$modes, $b_i$ and $b_j$, we consider the implementation of a $SWAP$ gate, cf. Fig.\ \ref{figswap} in Appendix \ref{quantumgates}. To achieve this, we need three (control) $c-$modes $c_1$, $c_2$ and $c_3$. Then the circuit is generated by the following sequence
\begin{equation}
H_b\to H_b + TOFF(c_1, b_i, b_j)\to H_b + TOFF(c_2, b_j, b_i)\to H_b + TOFF(c_3, b_i, b_j)\to H_b
\end{equation}
where $H_b=H_{b_i}+H_{b_j}$ and $TOFF(c_k, b_i, b_j)$ denotes the sequence of Hamiltonians that implements a Toffoli gate controlled by $c_k$ and acts on $b_i$ and $b_j$. Then, if $c_k=1$, a $CNOT$ is performed on $b_i$ and $b_j$ with $b_i$ being the control. Accordingly, this circuit acts on the modes as
\begin{equation}
|1 1 1\rangle_{c_1, c_2, c_3}\otimes |b_i b_j\rangle_{b_i, b_j} \to |1 1 1\rangle_{c_1, c_2, c_3}\otimes |b_j b_i\rangle_{b_i, b_j} 
\end{equation}
and, therefore, is equivalent to
\begin{equation}
U_{SWAP}|b_i b_j\rangle_{b_i, b_j}=|b_j b_i\rangle_{b_i, b_j} \hspace{10mm}\text{with}  \hspace{3mm} U_{SWAP}=\left(\begin{array}{cccc}
1 & 0 & 0 & 0      \\
0 & 0 & 1 & 0      \\
0 & 1 & 0 & 0       \\
0 & 0 & 0 & 1
\end{array}\right).
\end{equation}
The time required to implement this two-qubit gate is given by $t_{SWAP}\approx  3\,  t_{TOFF}$, where $t_{TOFF}$ is the time the circuit performing a Toffoli gate takes and is approximately given by $t_{TOFF}\approx 20 \,  t_{gate}^{BEC}$. Thus  $t_{SWAP}=\mathcal{O}(\epsilon^{-1})$ and a $SWAP$ gate  can be implemented in the same time as any other elementary gate.

\subsection{Universal quantum computing in black holes}\label{uqcbh}

To establish the universality of black hole type computers, we first characterize the available qubits. A general argument, based on the black hole entropy $S\sim N$, suggests that the number of qubits is given by $N$. The discrepancy between $N$ and the number of modes $N_{modes}$, we described above, most likely goes back to the simplicity of the toy model \cite{Dvali:2015ywa}. Then, for all states of the $N$ qubits to be sufficiently degenerate in order to account for the black hole entropy, the energy gaps $\epsilon_i$ of the qubits have to fulfill\footnote{As above, we work in units $\frac{\hbar}{R}=1$.}
\begin{equation}\label{energygaps}
\sum_i \epsilon_i \sim 1.
\end{equation}
That implies that for black holes almost all gaps $\epsilon_i$ scale as $\epsilon_i\sim\frac{1}{N}$ and the free Hamiltonian is given by
\begin{equation}
H=\sum_i \epsilon_i b_i^\dagger b_i.
\end{equation} 
In order to perform gate operations on these qubits, we include interactions to external $c-$modes described by the generating Hamiltonians of Sec.\ \ref{hamforgates}. As above, these Hamiltonians generate a universal set of quantum gates for the black hole qubits $b_i$. The existence of this set implies that black holes are universal for quantum computing. Due to the qubit energy gap $\epsilon_i$ and the gravitational interaction strength $\alpha\sim\frac{1}{N}$, the gate operations require the following gate operation time $t_{gate}$ 
\begin{equation}\label{timespecies}
t_{gate}\sim N.
\end{equation}
The existence of a minimal time per gate operation that scales with the universal black hole parameter $N$ has deep implications to the computational power of an evaporating black hole that has a life-time $t_{BH} \sim N$ which  of the same order as the Page time.

\section{Quantum computations}\label{seccomp}

In Sec.\ \ref{uqcbh}, we established that black hole type computer are universal for quantum computing. However, given the finite life-time $t_{BH}$ of an evaporating black hole that, in terms of $N$, is given by
\begin{equation}
t_{BH}\sim t_{Page}\sim N,
\end{equation}
the time that can be used to perform a computation is fundamentally finite. Therefore, the complexity of operations that can be implemented in a black hole type computer is restricted and, thus, despite the fact that there is a universal set of gates available, not all quantum circuits can be run on these systems. In the following, being mainly interested in the complexity of operations, we drop constant factors that to not affect the scaling properties of the relevant quantities. That also justifies to neglect the loss of qubits during the evaporation process. One can, for example, focus on half of the qubits  and limit oneself with respect to the available time by the Page time, as we will do. Then, although in practice the number of gates that can be applied is smaller as the system is shrinking during the evaporation process, it remains to be of the same order. 

At this point, we want to emphasize the key 	ingredients that go into the considerations of the present section
\begin{itemize}
\item[1.] the relevant degrees of freedom are well described by $N$ qubits, $\{b_i\}_{i=1, 2, \dots N}$, where $N$ equals the black hole entropy,
\item[2.] the life-time  is of the order  $t_{BH}\sim N$,
\item[3.]  the qubits have an energy gap $\epsilon$ of the order $N^{-1}$, 
\item[4.] there exists a universal set of gates acting on these qubits, where the time $t_{gate}$ to perform one gate operation is of the order of the inverse gap $\epsilon^{-1}$.
\end{itemize}
In the quantum $N-$portrait of black holes \cite{Dvali:2011aa, Dvali:2012en}, all these features are naturally present, as we discussed above in Sec.\ \ref{uqcinbogoinbh}. However, we emphasize that all following considerations in this section are widely model-independent.

\subsection{Fundamental limit on the circuit complexity}\label{fundlimit}

We estimate the maximal quantum circuit 	composed of the elements of our universal set of gates $\{ H, \,  R(\frac{\pi}{4}), \, CNOT\}$, obtained in Sec.\ \ref{hamforgates}, that can be implemented on qubits $\{b_i\}_{i=1, 2, \dots N}$.  The maximal number of gates that can be implemented in successive time-steps is given by the ratio of $t_{BH}$ and the time $t_{gate}$ required to apply one gate. However, gates can also be applied in parallel to different qubits $b_i$. This leads to the maximal number of gates in parallel given by the total number of qubits $N$. These considerations lead to a maximal circuit size (total number of gates) $S_{circuit}^{BH}$ and maximal circuit depth (time steps, or layers of gates) $C_{circuit}^{BH}$  given by
\begin{subequations}\label{circuitcplty}
\begin{align}
C_{circuit}^{BH}\sim& \frac{t_{BH}}{t_{gate}},\\
S_{circuit}^{BH}\sim& N\,\, C_{circuit}^{BH} \sim N \frac{t_{BH}}{t_{gate}},
\end{align}
\end{subequations}
where $t_{gate}$ is the time cost of a single gate operation. Interestingly, for black hole qubits, the gate operation time $t_{gate}$ is of the same order as the life-time of the black hole $t_{BH}$. In consequence, performing of the order $\mathcal{O}(1)$ gate operations per qubit is the maximum that can be achieved. Thus, we find
\begin{equation}\label{circuitspecies2}
C_{circuit}^{BH}\sim 1,\hspace{10mm}S_{circuit}^{BH}\sim N.
\end{equation}
We want to emphasize that the circuit size complexity $S_{circuit}^{BH}$ is completely characterized by the universal number $N$. Furthermore, the triviality of the circuit depth $C_{circuit}^{BH}$ is a remarkable fact. The interpretation of bounds (\ref{circuitspecies2}) is the following. Due to the  gravitational coupling in black holes the gate operation time and the black hole life-time are nearly identical. Thus, only a constant number of gates can be applied in sequence. In consequence, the maximal circuit depth is trivial. These features make the black hole type computer a poor quantum computer. Even so it is universal, the complexity of computations that can be carried out using it is very limited. Therefore, in a strict sense, it is not a universal quantum computer. However, nobody would expect the existence of a truly universal quantum computer in nature anyway. In the following, we discuss the issue of decoherence and its effects.

\subsection{Local decoherence and entanglement} \label{decoherencesec}

As pointed out in  \cite{Dvali:2012en, Dvali:2015ywa, Dvali:2015wca, Dvali:2016wca}, one of the interesting properties, from a quantum information point of view,  of qubits appearing near the quantum critical point is their weak coupling 
to the rest of the system.   This is important for the longevity of information storage in a given qubit.  In general,  after some time the system evolves into a state in which a given qubit becomes entangled with the rest of the
system. Information that was originally stored in the qubit then gets 
scrambled among the rest of the qubits.  From the point of view of an observer that wants to perform a quantum computation with a given qubit, but is 
blind with respect to the rest of the system,  such entanglement will look as an effective decoherence. We shall refer to this effect as {\it local decoherence}  or {\it maximal self-entanglement}.  
 
 However, the weakness of the coupling can make such a decoherence time long.
We shall now discuss this point. We shall try to keep the discussion maximally 
general and model-independent. 
Let us consider the system of $N_q$ qubits $b_j, ~j = 1, 2,\dots, N_q$.  
The Hilbert space of the system can be described by a $2^{N_q}$-dimensional Fock space with basis vectors 
 $ |n_1, n_2, \dots, n_{N_q} \rangle \equiv  |n_1\rangle \otimes  |n_2\rangle \otimes\dots \otimes |n_{N_q} \rangle$  where $n_j =0,1$ are the eigenvalues of 
 $b_j^{+} b_j$.   Of course,  the choice of the basis is a matter of convenience and one can pick up
 any other possibility.   Let the initial state vector of the system be given by a tensor product state 
 $|\psi_1\rangle \otimes  |\psi_2\rangle \otimes \dots \otimes |\psi_{N_q} \rangle$, where 
 $|\psi_j\rangle = \alpha_j |0\rangle +  \beta_j |1\rangle$ with $|\alpha_j|^2   +  |\beta_j|^2 = 1$.  If qubits are decoupled from each other, the state will evolve as a tensor product.  In such a case an external observer 
 can manufacture  a logic gate by coupling some external mode $c$ to one of the qubits, e.g., in one of the ways discussed in Sec.\ \ref{hamforgates}, and can perform logical operations over it without worrying about the rest of the qubits.   

  However, since 
 the qubits are coupled to each other, the state $|\psi_1\rangle \otimes  |\psi_2\rangle \otimes \dots \otimes |\psi_{N_q} \rangle$,  will evolve into an entangled state.  A maximal local decoherence (i.e., maximal self-entanglement) will take place after the state will evolve into a generic superposition of  $2^{N_q}$ basis vectors.
At this point, an observer that wants to perform a measurement on any given qubit, $b_k$,   will not be able 
to even approximately treat the state as a tensor product of the type  $|\psi_k\rangle \otimes |\psi_{rest} \rangle$,
where $|\psi_{rest} \rangle $ is a state describing the rest of the system.  Of course, the argument is probabilistic
and there is a non-vanishing but small probability for the system to evolve into an atypical maximally-entangled state 
with smaller number of basis vectors involved. An extreme example would be a state, 
 $ {1\over \sqrt{2}} \left (  |0\rangle_{1} \otimes  |0\rangle_{2} \otimes \dots \otimes |0 \rangle_{N_q}  \,  + \, 
   |1\rangle_{1} \otimes  |1\rangle_{2} \otimes \dots \otimes |1 \rangle_{N_q} \right ) $, which is maximally entangled despite the fact that only two basis vectors are involved in the superposition.  

 We need to estimate how long it takes for the system of $N_q$ qubits 
in order to evolve into a typical maximally-entangled state
that would involve a superposition of order  $2^{N_q}$ basis vectors.
 We shall call this time $t_{decoh}$.  The  physical meaning of this time-scale  is that 
 the information stored in any given single qubit becomes maximally 
 distributed among all $N_q$ qubits.  
  We wish to show that  a model-independent bound on this time-scale can be established solely from the knowledge of the entropy of the system and
the assumption that the qubits contributing in it are similar.  
That is,  none of the  $N_q$  qubits that contribute into the entropy-counting must be  privileged  in any way with respect to others.     

   Obviously, $t_{decoh}$  is determined by the strength of the off-diagonal terms in the Hamiltonian that mix different qubits, e.g., 
  \begin{equation}
   H_{int} \, = \,  \sum_{ij} \,  \alpha_{pairs}^{ij} \,  b_i^+b_j^+ b_ib_j \, + \dots\, ,
  \label{intHb} 
 \end{equation}      
 where $\alpha_{pairs}^{ij}$  are parameters.  
    This strength can be estimated from the knowledge of the entropy assuming a certain level of democracy (no privilege) among the qubits contributing into the entropy. 
   Namely, as said above, we shall assume that the $N_{q}$ qubits   
 are similar. This means,  for example, that the number of partner 
   qubits, $N_{partner}^{(j)}$,  to which a given qubit  $b_j$ mixes in the Hamiltonian, is approximately universal. That is,  to leading order, $N_{partner}^{(j)}$ is $j$-independent.  Thus, all qubits  have roughly equal number of partners given by an universal number $N_{partner}$. Obviously, this number satisfies the condition $0 < N_{partner} < N_q $.  
    
    Similarity of qubits also implies that the strength of the mixing among the pairs of partner qubits  in the Hamiltonian is approximately universal and is given by a certain parameter $\alpha_{pairs}$.  
  Notice, since in the case of the critical system qubits are {\it collective}  excitations of some ``original" degrees of freedom, the coupling  $\alpha_{pairs}$ in general is different from the coupling $\alpha$ of these original degrees of freedom. This is, for example, clear by comparing the coupling between the $a$-modes  
in (\ref{Haa}),  which represent bosons that  form the critical Bose-Einstein condensate, and the  
coupling (\ref{lagbogo2}) between the Bogoliubov $b$-modes that represent collective excitations of $a$-modes. 

   Furthermore, $2^{N_q}$ states populating the Fock space of  $|n_1, n_2, \dots n_{N_q} \rangle$  fall within the energy gap $\sim N_q \epsilon $.   
 The  contribution to the energy from the interaction terms among the qubit pairs, when evaluated over the states in which order-$N_q$ qubits are 
in excited states,  will be  given by the number of interacting pairs $N_{pairs}$
  times the coupling $\alpha_{pairs}$.  Thus, the maximal  average contribution to the energy from the 
  off-diagonal terms in the Hamiltonian can  be estimated as $
     E_{int}  \sim N_{pairs} \alpha_{pairs}$.
    Since each given qubit couples to  $N_{partners}$ partners, the total number of pairs (neglecting 
   numerical factors order one) is 
    $N_{pairs} \sim N_q N_{partners}$.  This gives us an estimate of an upper bound on the interaction energy over the states of interest,  
     \begin{equation}
     E_{int}  \sim N_q N_{partners} \alpha_{pairs} \, .
     \label{interactionE}
   \end{equation}          
      Since the above-considered $N_q$ qubits contribute into the entropy of the system with $2^{N_q}$ micro-states fitting within the 
energy gap $N_q\epsilon$, we arrive at the  following constraint. This constraint comes from the obvious requirement that the  interaction energy should be below the total energy gap which houses the states contributing into the entropy, i.e.,  $ E_{int}  \lesssim N_q\epsilon$. Taking into account 
 (\ref{interactionE}) this gives, 
    \begin{equation}
    N_{partners} \alpha_{pairs} \lesssim \epsilon \, .
     \label{const1}
   \end{equation} 
   Notice, the left hand side of this equation measures the maximal off-diagonal    disturbance that  can be exerted on each individual qubit from its partners.  Thus, according to (\ref{const1}) the off-diagonal disturbance is always below 
 the diagonal value of the Hamiltonian, which is given by $\epsilon$.   Thus, the energy is dominated by diagonal terms in the Hamiltonian.  In some sense this is not surprising, since in the opposite case the set of   $N_q$ qubits will not be a reliable counter of the entropy and this would contradict to the starting assumption.

  Notice, in the case when qubits  $b_j$ originate from the underlying quantum  criticality of the system, as e.g., it is  the case for (\ref{lagbogo2}),  the bound (\ref{const1}) can be understood as a consequence of quantum criticality:  off-diagonal terms in the Hamiltonian cannot offset the quantum criticality of the system and thus must contribute less than the diagonal terms.  In this sense,  assumption of criticality substitutes the need for the assumption that qubits are reliable entropy contributors and vice-versa. 

  As we saw, with the latter assumption, the bound (\ref{const1}) was derived without any reference to a particular origin of qubits and therefore is much more general.  
 As long as the entropy of the system results from the micro-states 
 produced by set of similar qubits of gap $\epsilon$, the bound follows. Bound (\ref{const1}) immediately  translates into  a bound on 
   $t_{decoh}$. 
 Indeed,  the off-diagonal influence from the mixing with its partners is what  measures the efficiency of a given qubit to entangle with the rest of the system.  
 The corresponding time scale, $t_{decoh}$  is set by the inverse value of the corresponding  off-diagonal mixing, i.e., with the left hand side of  (\ref{const1}).

   We thus arrive at a very general and powerful bound. 
  Irrespectively how the qubits are grouped in the sets of interacting pairs, the local decoherence time is universally-bounded from below by the qubit energy gap
  \begin{equation} 
  t_{decoh} \gtrsim \epsilon^{-1}  \, .   
 \label{decoherence} 
 \end{equation}    
 
  The physical intuition behind this bound is very transparent and can be summarized in two sentences. Indeed, in order for a given qubit energy gap, 
 $\epsilon$, not be disturbed by its partners, the off-diagonal mixing terms with the partners in the Hamiltonian must be small compared to the diagonal contribution coming from the qubit gap.  
 On the other hand the minimal time-scale for a qubit state to be influenced
by its partners (i.e., to become fully entangled with them)  is precisely given by the inverse value of this off-diagonal Hamiltonian.  Hence the bound 
 (\ref{decoherence}).

  This simple bound has a deep implication when applied to black holes, because 
  it shows that for black holes the local decoherence time (i.e., the time during which black hole qubits fully entangle among each other),   
 is not shorter than its 
  half-life time.  Of course, to make the bound more concrete we must know  the two quantities  
$N_{partners}$ and  $\alpha_{pairs}$.  This information about the black holes 
we do not posses, but this also is irrelevant.  Since for the black hole 
$\epsilon^{-1} \sim N$ is of the order of black hole half life time and we know that at least half of the black hole qubits are strongly affected by this time due to Hawking evaporation.   Putting it differently, black hole qubits locally-decohere  
only after half of the black hole is gone.

\subsection{Connection to scrambling} \label{scramblingsec}

We showed in the previous section that the minimal time scale $t_{decoh}$, during which 
the black hole qubits become fully entangled with each other, is comparable to the black hole half-life time. 
Obviously, $t_{decoh}$  is the scale during which the message is maximally distributed among all the qubits, and therefore we can say that the message is 
{\it maximally scrambled}. We have to be very clear with the use of this term, 
  since in the literature a much shorter time-scale has also 
  been referred to as the scrambling time. 
  This time-scale,  which originally has been suggested by Hayden and Preskill in 
  \cite{Hayden:2007cs} as the scrambling time for a black hole,  
 is much shorter and (in our units) scales as $t_{scramb} \sim ln(N)$.  Moreover an explicit 
 microscopic confirmation to the existence of this time scale was given in 
 \cite{Dvali:2013vxa}, where it was shown that the critical Bose-Einstein model 
 given by (\ref{hamiltonianring}) indeed exhibits a scrambling time 
 $t_{scramb} \sim ln(N)$.  Namely, in this paper the connection between the fast scrambling and the existence of the Lyapunov 
 exponent and chaos was established.  Since to our knowledge
 the critical Bose-Einstein condensates are the only existing microscopic models in which the origin of both 
 time-scales has been explicitly traced, we shall try to use this knowledge  
 for distinguishing the meanings for the two scales from quantum information perspective.  As in the previous section, we would like to make the 
 arguments maximally model-independent.  
 
 As we shall explain in very general terms, both scales can coexist 
 and be given the meaning of information scrambling, but to different extends.   In order to see this,  let us perform the following thought experiment. 
     Let  us assume that the initial state of a black hole is given by 
   a single basis vector in the Fock space of $N$-qubits, which for definiteness we chose to be  $ |\psi \rangle_{t=0} \,  = \, |0, 0, \dots, 0 \rangle $.   
  As we discussed in the previous section, the minimal time-scale it takes  this state to evolve into a typical superposition of order-$2^N$ basis vectors 
  is $t_{decoh} \sim N$. If we naively extrapolate this dynamics to shorter time-scales, we can model the growth of the number of basis vectors entering in the superposition representing the full state vector,  by an exponential law, 
  $n_{members}(t) \sim 2^t$. With this modeling after $t = ln(N)$ time  
  the initial basis vector will evolve into a superposition of 
  $\sim N$ basis vectors. If we combine this fact with the knowledge that the off-diagonal terms in the Hamiltonian for each qubit, according to (\ref{const1}),  are at most $\sim 1/N$, we arrive to the following estimate. 
  After  $t=ln(N)$ the state vector represents a superposition of   $\sim N$ basis vectors,  such that in this superposition the eigenvalue of each qubit is 
 flipped (relative to its initial value) with probability $1/N$. 
   Thus,  after $t=ln(N)$ the initial state will evolve - with high probability - into a superposition of $N$ basic vectors  in which each eigenvalue is altered 
 (relative to its initial value) roughly in one member of the superposition only.
  That is,  the initial vector  is expected to evolve into something of the following sort, 
   \begin{equation} 
  |\psi \rangle_{t=ln(N)} \,  = \, {1\over \sqrt{N}} \left(  |1, 0, \dots, 0 \rangle \, + \,  |0, 1, \dots, 0 \rangle \, +\dots+ \,  |0, 0, \dots, 1 \rangle \right)\, .  
  \label{aftertime}  
   \end{equation}  
   In the above expression, order one coefficients  and phase factors 
   are ignored.   
  Now imagine an observer (Alice) that works with a given qubit of a black hole, say 
  $b_1$ and ignores the rest.   
    We shall assume that Alice can perform a measurement over a state vector and we shall ignore the technicality that such measurement can take a very long time. Now let us imagine that Alice encoded a message in the initial state vector of a black hole, 
    but in a particular qubit  $b_1$.  That is, she gives a meaning of a message to the state $|0\rangle_1$.
  
   What happens with this  message after the time $t=ln(N)$? 
   The message becomes distributed among $N$ states: state vector is superposition  in which all the qubits participate democratically,   
   each with $1/N$ probability.  Therefore, for an observer that does not know 
   the identity of the original qubit of Alice, it would be natural to say that the message 
   got scrambled. 
    However,  for the observer that knows the identity of the original qubit, the story is different. For Alice, since she only cares about the qubit $b_1$, the state  remains almost intact, since the state can still be represented as a  tensor product  of the state of $b_1$-qubit with the rest up to ${1\over \sqrt{N}}$ admixture, 
       \begin{equation} 
  |\psi \rangle_{t=ln(N)} \, = \,  |0\rangle_1 \otimes |rest \rangle \, +  {1\over \sqrt{N}} |1 \rangle_1\otimes | 0\rangle_2 \otimes \dots\otimes |0 \rangle_N \,,  
  \label{aftertime1}  
   \end{equation}    
 where $|rest \rangle$ only includes the sates of $b_{j\neq 1}$.   
 From above, it follows that we can refer to the above time-scale as the time 
 of {\it minimal scrambling}, $t_{min.\,scramb} \sim ln(N)$. 
 
  Notice the crucial difference, between the time $t_{min.\,scramb}$  and the 
  time  $t_{decoh}$:    after $t_{decoh}$ - since the state vector evolves into a generic  superposition of all the $2^N$ basis states -  even the observer that can trace a given qubit can no longer recover 
    the message that was initially stored in it without knowing the states of other qubits.  From this point of view we can refer to $t_{decoh}$ as the time of {\it total scrambling}.    
    
    Of course, under no circumstances the scrambling of a message must be confused with the loss of information. The states of the system we consider remain pure all the time and the evolution is unitary.  Correspondingly, the entire information about the state vector is in principle accessible. 
    In our description, scrambling is only a measure of difficulty of this recovery at  
 various stages of the system evolution.  In the regimes of our interest, although 
 after $t=ln(N)$ the message gets scrambled in a certain minimal sense,
 for an observer that is performing a quantum computation with a given qubit this 
 poses no difficulty. For such a observer, the problems appear after the time 
 $t_{decoh} \sim N$, but by this time other more severe problems set in, because  half of the black hole is simply gone into Hawking radiation.

 \subsection{Efficiency of computations}\label{efficiency}

After having discussed the internal interactions that lead to local decoherence, we now study the efficiency of computations. In \cite{RIS}, given a fixed amount of energy $E$, a bound on the speed of computations in terms of operations per time was established.\footnote{A related but slightly different bound is considered in \cite{Brown:2015bva, Brown:2015lvg}. Here, however, we restrict to computations that proceed in steps that are given by mutually orthogonal states and therefore use Lloyd's bound \cite{RIS}.} It is given by
\begin{equation}\label{llodybound}
\sum_{gates} \frac{1}{t_{gate}} \leq \sum_{gates} E_{gate}= E,
\end{equation}
where $E_{gate}$ is the energy required to perform a gate operation, $t_{gate}$ is the gate operation time and we dropped constants of order $\mathcal{O}(1)$. Saturation of bound (\ref{llodybound}) implies a maximally fast computation that we refer to as being maximally efficient. Here, the energy involved in a computation is $E=N \epsilon\sim 1$ and the gate time is $t_{gate}\sim N$. Thus, bound (\ref{llodybound}) is saturated and we can conclude that black hole gates work at the maximal speed. However, the energy required to build a black hole, $E=M\sim N$, is much larger than $N \epsilon\sim 1$. Thus, the speed of computations in a system of energy $E\sim N$ can, in principle, be much higher,
\begin{equation}
\sum_{gates} \frac{1}{t_{gate}}\sim  N.
\end{equation}
That is, even so the gates work at the maximal speed, given by the energy $E=N \epsilon\sim 1$, we have to conclude that it is a slow computer compared to a system of energy $E\sim N$ that could potentially perform computations maximally fast.\footnote{Let us mention that it is not clear that such a system exists.} That is, because the systems we are considering here, are only using a tiny fraction of their energy on performing gate operations, while most of the energy is spend on maintaining criticality. However, the part of their energy that is available for computations is used maximally efficient. To summarize, the quantum gates are maximally efficient due to the criticality of the system. It is also this critical behavior that allows for a huge entropy, i.e., a huge number of qubits. Thus criticality provides an enormous memory. However, only a  fraction $\frac{1}{N}$ of the total energy is available for computations, while most energy is radiated away in form of Hawking radiation during the operation time of the gates.

\subsection{Resolution of the micro-state and retrieval of information}\label{aspects}

To discuss a possible resolution of the micro-state of the system, we first characterize the class of unitaries that can be implemented  according to the micro-states that can be reached by these operators starting from  a product state. Every state $|\psi\rangle$ in the Hilbert space $\mathcal{H}$ of micro-states  can be associated with a unitary operator $U_\psi$ according to $|\psi\rangle=U_\psi |BH\rangle$ for a fixed reference state $|BH\rangle$. Here, for simplicity, we start from a black hole based computer with all constituents being in a product state
\begin{equation}\label{bhprodstate}
|BH\rangle=\bigotimes_{i=1}^{N}|b_i\rangle
\end{equation}
for some fixed $b_i$'s and neglect internal dynamics in the following. The main results we obtain would not be altered by skipping these simplifying assumptions. The subsets of states $\{\psi_j\}$ in the black hole Hilbert space can be classified according to the complexity of the unitary matrix $U_\psi$ that generates the respective state by acting on the reference state. Given the triviality of (\ref{circuitspecies2}), we restrict ourselves to the subset of states that are associated with unitaries in the trivial sector, i.e., with states $|\psi_j\rangle$ that are generated by the action of a unitary  of trivial circuit depth $C_{circuit}(U_{\psi_j})\sim 1$. It is clear that these are the states $|\psi_j\rangle$ that can potentially be reached by a quantum circuit. However, the details depend on the exact value of the number of gates that can be applied successively. To be precise, a state $|\psi_j\rangle$ is reachable if
\begin{equation}\label{condreadout}
C_{circuit}(U_{\psi_j})\leq  C_{circuit}^{BH}.
\end{equation}
In consequence, only a small fraction of the total Hilbert space $\mathcal{H}$ can be explored. However, the set of micro-states that can be partially resolved\footnote{We define (partial) resolution as the transfer of the product state of a (sub-)system to external modes.} is larger.  Consider, for example, a state of the form
\begin{equation}
|\psi_j\rangle=U^{(n_1)}\otimes U^{(n_2)} \otimes \dots\otimes U^{(n_m)}|BH\rangle,
\end{equation}
where each of the $U^{(n_i)}$ acts on $n_i$ qubits and  some of the $U^{(n_i)}$ have a  non-trivial circuit complexity. Then, there is some subset of qubits in state
\begin{equation}
U^{(n_i)} \bigotimes_{k=j}^{j+n_i-1} |b_k\rangle
\end{equation}
that is separable from the state of the remaining qubits. In this case a resolution of the state is possible if a modified version of (\ref{condreadout}) holds, i.e., if $C_{circuit}(U^{(n_i) ^{-1}})+1\leq  C_{circuit}^{BH}$. Finally, there is the special case of having some qubits $b_i$ in a product state. This is the case, when $U^{(n_i)}=\mathbb{1}^{(n_i)}$. These can be resolved if $C_{circuit}^{BH}\geq 1$. These results show that, even so some states can potentially be resolved, most of the Hilbert space $\mathcal{H}$ remains hidden due to the limitation given by (\ref{circuitspecies2}).

The next point we want to address is a rough comparison with information retrieval by Hawking radiation. Here we define information retrieval as an operation $O$ achieving $O |\psi\rangle=|BH\rangle$. Note that this definition is not unique, as it depends on the reference state $|BH\rangle$. The idea behind this definition is that $U_{\psi}$ carries all the information and inscribes this information into $|BH\rangle$ by producing $|\psi\rangle$.  We note that the action $O |\psi\rangle=|BH\rangle$ can equally well be viewed as $|\psi\rangle=O^\dagger|BH\rangle$ for unitary $O$. Therefore, according to the above definition, information retrieval is equivalent to encoding. For example, given the reference state $|BH\rangle=\bigotimes_{i=1}^{N}|0\rangle$ and the state of interest $|\psi\rangle=\bigotimes_{i=1}^{N-1}|0\rangle\otimes |b_i \oplus 1\rangle$, the information is retrieved by $O=\mathbb{1}\otimes\dots\otimes \mathbb{1}\otimes \sigma_x$, where $\sigma_x$ is the flip operator. Alternatively, the information is inscribed as $O^\dagger|BH\rangle=|\psi\rangle$.

Let us consider the system being in the state $|\psi_{BH}\rangle=U_{\psi_{BH}}|BH\rangle$.\footnote{Here, we only consider pure states. However, the present conclusions carry over to mixed states straight forwardly.} Then a crude way to extract the state is by swapping the state of each qubit (or subsystem of qubits) with a (separate) external mode $c_i$, as 
\begin{equation}
|\psi_{BH}\rangle_{b_1,b_2,\dots,b_{N_{}}} \otimes |\psi_{ext}\rangle_{c_1,c_2,\dots,c_{N_{}}}\overset{U}{\longrightarrow} |\psi_{ext}\rangle_{b_1,b_2,\dots,b_{N_{}}} \otimes |\psi_{BH}\rangle_{c_1,c_2,\dots,c_{N_{}}}, 
\end{equation}
where $|\psi_{ext}\rangle_{c_1,c_2,\dots,c_{N_{}}}$ denotes some product state of the external $c-$modes. The unitary $U$ that swaps the state of the black hole with the state of the $c-$modes is given by
\begin{equation}\label{uniswap}
U=SWAP(b_1,c_1)\otimes SWAP(b_2,c_2)\otimes\dots\otimes SWAP(b_{N_{}},c_{N_{}}),
\end{equation} 
where $SWAP(b_i,c_i)$ is the $SWAP$ gate applied to the $i$th $b-$ and $c-$mode. In this way the state of the black hole is ``transferred'' to the external $c-$modes by a circuit of size $\mathcal{O}({N_{}})$ and depth $\mathcal{O}(1)$. In consequence, any information present in the black hole is available for decoding after the gate operation time $t_{gate}\sim N$. This can, in principle, be a message that fell into the black hole or simply the information about the state of the collapsing matter forming the black hole.

If instead one decides to gather the Hawking radiation to obtain this information, one ends up, after the black hole evaporation time $t_{BH}\sim N$, with the radiation in the  state $|\psi_{rad}\rangle=U_{ev} U_{\psi_{BH}}|BH\rangle$, with $U_{ev}$ being a unitary describing the dynamics of the evaporation process. Making the assumption that the evaporation process is local, it is to be expected that $C_{circuit}(U_{ev})$ is maximally of polynomial complexity. Therefore, given that the black hole state is of polynomial or of exponential complexity, $C_{circuit}(U_{\psi_{BH}})=\mathcal{O}(\text{poly}(N))$ or $C_{circuit}(U_{\psi_{BH}})=\mathcal{O}(\text{exp}(N))$, the state of the radiation is of the same complexity. That is
\begin{equation}
C_{circuit}(U_{ev} U_{\psi_{BH}})\sim C_{circuit}(U_{\psi_{BH}}),
\end{equation}
where, here, $``\sim"$ indicates the same complexity class (polynomial or exponential). So far we did not specify the computational capabilities of the external observer, Alice, that either couples to the black hole qubits or collects the Hawking radiation. However, it seems reasonable to assume that her computational power is such that, given a (non-trivial) unitary $U_{\psi_{BH}}$, there is essentially no difference between the time required to decode either $\psi_{BH}$ or $\psi_{rad}$. Therefore, the information in the state $\psi_{BH}$ cannot be accessed faster by coupling directly to the black hole qubits than by collecting and decoding the emitted Hawking radiation. At this point we want to note that the inclusion of the effects discussed in Secs.\ \ref{decoherencesec} and \ref{scramblingsec} does not alter this conclusion.

In summary, the bound (\ref{circuitspecies2}) implies that only a tiny fraction of the Hilbert space $\mathcal{H}$ can be explored. Furthermore, considering a state $\psi_{BH}$ of non-trivial complexity, information cannot be retrieved directly by means of externally coupling to the black hole. The reason is that all unitary operations that can be implemented are restricted to be accomplished by a quantum circuit of trivial depth, as the gate operation time $t_{gate}$ and the life-time $t_{BH}$ of the black hole are approximately equal, $t_{gate}\sim t_{BH}$. In consequence, an operation that transfers the state of a qubit to an external system takes approximately the same time, as it would take waiting outside the black hole for the respective Hawking radiation to be emitted. Therefore, one cannot acquire knowledge about the black hole state before the information naturally left the black hole in form of radiation. However, due to the fact that it might be possible to resolve part of the black hole micro-state by coupling to the internal degrees of freedom, they act as hair for the black hole. Interestingly, in the classical limit, $N\to\infty$, these qubits become (very low-energy) excitations similar to the soft hair related to supertranslations in \cite{Averin:2016ybl}. On the classical level these soft hair are zero-energy excitations \cite{Hawking:2016msc, Dvali:2015rea}.

\section{Conclusions}\label{conclusions} 

The recently established isomorphy between the well-known quantum information properties of black holes and attractive  Bose-Einstein condensates at quantum criticality allows us to analyze the quantum computational properties of black holes in the language of simple prototype Bose-Einstein systems. 
  
 In this language the degrees of freedom that store and process quantum information are identified with collective Bogoliubov modes of  the condensate. The vanishing energy gap and extremely weak coupling of Bogoliubov qubits make them into viable candidates for such a role.    Exploiting this connection, the idea of performing a black hole based quantum computational sequence  in critical Bose-Einstein systems was put forward  in \cite{Dvali:2015wca}.  It was shown that by coupling to external modes one can design logic gates and perform quantum computations while maintaining criticality of the system.   It is apparent that in such a case the minimal time-scale of a logical operation is given by the inverse gap at criticality $\epsilon^{-1}$. 

   In the present paper we have extended this analysis and addressed the questions of complexity, universality and efficiency of such a computation.    As in the previous case,  we have studied the     system from the point of view of an observer that  can perform computations via Bogoliubov qubits by coupling  them to  external degrees of freedom. The requirement is that the external influence must be soft-enough in order not to disturb  the quantum criticality   of the system.  We refer to the computers designed in this way as {\it black hole based quantum computers}.

 We have established that such black hole based computers are universal for quantum computing, and found that the complexity of circuits running on these systems is trivial.  We thus also concluded that atomic Bose-Einstein condensates with attractive interactions can be employed as universal quantum computers.  Furthermore, we outlined concrete approaches to gain partial knowledge about the micro-state. Finally, we studied the retrieval of information and found that the advantage one gains by externally coupling to a black hole is negligible compared to the gathering of the emitted Hawking radiation. Our findings show, in accordance with \cite{Page:1993wv}, that Hawking radiation is the key for the retrieval of information, as a direct readout is not possible. The reason for that is given by the trivial circuit depth of any quantum circuit that may be implemented.
 
   The system of $N_q$ qubits  is characterized by some important time scales.  One of these, $t_{decoh}$,   is the time that it takes for a given qubit to maximally entangle with the rest of the qubits.  After this time, a message stored in the state of the original qubit becomes totally scrambled.  Thus, the time of  local decoherence  $t_{decoh}$ simultaneously represents the time of maximal entanglement as well as the time of {\it total scrambling}.   We have established a model-independent bound on the time-scale $t_{decoh}$  given by (\ref{decoherence}).   We have clarified the physical meaning   - within the given framework - of the minimal scrambling time $t_{min.\,scamb.}$.     The latter  is the minimal time-scale that it takes a message - initially stored in an atypical  state consisting of a single basis vector - to be diversified in a ``democratic" superposition  of order-$N_q$ basis vectors obtained from the original state by altering the eigenvalue of each qubit  approximately once.  We called this time minimal scrambling due to the fact that, although the message is democratically redistributed, nevertheless an observer that can trace the qubit in which the message was originally stored, can detect it up to an order-$1/N$ error.    This is no longer true after the time $t_{decoh}$, after which  the system evolves   into a state in which all $N_q$ qubits are maximally-entangled and even an observer that can trace the state of the original qubit no longer is able to read out the message without knowledge of the state of all the qubits.

 When applied to the specific case of black holes or critical condensates the different  time-scales become related via (\ref{ttscales}).  It thus becomes apparent that the universal gravitational coupling in black holes that allows these systems to store an incredible amount of information for a macroscopically long time is responsible for the poor information processing capacity. That is, the gravitational coupling, that is responsible for the criticality of a black hole and sets the gap of the qubits, also gives the interaction strength for the coupling to the external modes. Therefore, the black hole life-time and the gate operation time have the same scaling properties. Despite this fact, we found that the gates satisfy Lloyd's bound for the gate operation time and are, in this sense, maximally efficient. However, due to the fact that only a tiny fraction of the system's energy is available for computations their capabilities are very limited. In fact, almost all energy is used for the maintenance of criticality that requires the black hole to evaporate due to Hawking radiation. Furthermore, the possibility of partial resolution of the  micro-state gives hair to the black hole.  In agreement with previous results, these hair vanish in the classical limit, as the qubit gap closes. 

 Finally, along  the lines  of \cite{Dvali:2015wca},  it would be important to capitalize on   the fact that, unlike the case of astrophysical black holes,  more  flexibility of manipulations with critical condensates can potentially allow us to combine the useful features of black hole type quantum computing, such as, the cheap information storage and macroscopically-long  local decoherence time, with the possibility of faster information-processing,  by coupling with external  modes that can take system in and out of criticality.

\tocless{\begin{acknowledgements}

The work of G.D. was supported by Humboldt Foundation under Alexander von Humboldt Professorship,  by European Commission  under ERC Advanced Grant 339169 ``Selfcompletion'' and  by TRR 33 ``The DarkUniverse''. The work of C.G. was supported in part by Humboldt Foundation and by Grants: FPA 2009-07908, CPAN (CSD2007-00042) and by the ERC Advanced Grant 339169 ``Selfcompletion'' . The work of D.L. was supported by  the ERC Advanced Grant 32004 ``Strings and Gravity" and also by TRR 33. The work of  Y.O. and B.R. was supported by Funda\c{c}\~{a}o para a Ci\^{e}ncia e a Tecnologia (Portugal), namely through programmes PTDC/POPH/POCH and projects UID/EEA/ 50008/2013, IT/QuSim, partially funded by EU FEDER, and from the EU FP7 project PAPETS (GA 323901). B.R. acknowledges the support from the DP-PMI and FCT through scholarship SFRH/BD/52651/2014. Furthermore, B.R. would like to thank the Arnold Sommerfeld Center for Theoretical Physics at LMU Munich for hospitality and support.

\end{acknowledgements}}

\appendix

\section{Quantum gates}

\subsection{Elementary quantum gates}\label{quantumgates}

\begin{figure}
 \centering
 
\begin{subfigure}{0.9\textwidth}

\begin{tikzpicture}[thick]

    \tikzstyle{operator} = [draw,fill=white,minimum size=1.5em] 
    \tikzstyle{phase} = [draw,fill,shape=circle,minimum size=5pt,inner sep=0pt]
    \tikzstyle{surround} = [fill=green!05,thick, draw=black,rounded corners=2mm]
    \tikzstyle{cnot} = [draw,shape=circle,minimum size=5pt,inner sep=0pt]

  \tikzset{
operator/.style = {draw,fill=white,minimum size=1.5em},
operator2/.style = {draw,fill=white,minimum height=3cm},
phase/.style = {draw,fill,shape=circle,minimum size=5pt,inner sep=0pt},
surround/.style = {fill=green!05,thick,draw=black,rounded corners=2mm},
cross/.style={path picture={ 
\draw[thick,black](path picture bounding box.north) -- (path picture bounding box.south) (path picture bounding box.west) -- (path picture bounding box.east);
}},
crossx/.style={path picture={ 
\draw[thick,black,inner sep=0pt]
(path picture bounding box.south east) -- (path picture bounding box.north west) (path picture bounding box.south west) -- (path picture bounding box.north east);
}},
circlewc/.style={draw,circle,cross,minimum width=0.3 cm},
}

    \matrix[row sep=1.2cm, column sep=1.6cm,] (circuit) {
    
    \node (q1) {$|a\rangle$}; &[-0.5cm] 
    \node[phase] (P11) {}; &
    \node[circlewc] (P12) {}; &
    \node[phase] (P13) {};  &
    \node (P14) {$|b\rangle$}; &[0.5cm] 
    \coordinate (end1); \\

    \node (q2) {$|b\rangle$}; &[-0.5cm] 
    \node[circlewc] (P21) {}; &
    \node[phase] (P22) {}; &
    \node[circlewc] (P23) {}; &
    \node (P24) {$|a\rangle$}; &[0.5cm] 
    \coordinate (end2);\\
   };

   \draw[thick] (q1) -- (P14)  (q2) -- (P24)  (P12) -- (P22)  (P11) -- (P21)  (P13) -- (P23);

    \begin{pgfonlayer}{background}
      
       \node[surround] (background) [fit = (q1) (P21) (P24)] {};

    \end{pgfonlayer}
    \end{tikzpicture}
\caption{A $SWAP$ gate can be implemented using three $CNOT$ gates. The states of the two input qubits are exchanged due to the subsequent application of three $CNOT$ gates.} \label{figswap}

\end{subfigure}

\vspace{4mm}

\begin{subfigure}{0.9\textwidth}

\begin{tikzpicture}[thick]

    \tikzstyle{operator} = [draw,fill=white,minimum size=1.5em] 
    \tikzstyle{phase} = [draw,fill,shape=circle,minimum size=5pt,inner sep=0pt]
    \tikzstyle{surround} = [fill=green!05,thick, draw=black,rounded corners=2mm]
    \tikzstyle{cnot} = [draw,shape=circle,minimum size=5pt,inner sep=0pt]

  \tikzset{
operator/.style = {draw,fill=white,minimum size=1.5em},
operator2/.style = {draw,fill=white,minimum height=3cm},
phase/.style = {draw,fill,shape=circle,minimum size=5pt,inner sep=0pt},
surround/.style = {fill=green!05,thick,draw=black,rounded corners=2mm},
cross/.style={path picture={ 
\draw[thick,black](path picture bounding box.north) -- (path picture bounding box.south) (path picture bounding box.west) -- (path picture bounding box.east);
}},
crossx/.style={path picture={ 
\draw[thick,black,inner sep=0pt]
(path picture bounding box.south east) -- (path picture bounding box.north west) (path picture bounding box.south west) -- (path picture bounding box.north east);
}},
circlewc/.style={draw,circle,cross,minimum width=0.3 cm},
}

    \matrix[row sep=1cm, column sep=0.8cm,] (circuit) {
    
    \node (q1) {$|a\rangle$}; &[-0.5cm] 
    \node[crossx] (P11) {}; &
    \node[phase] (P12) {}; &
    \node[crossx] (P13) {};  &
        \node[phase] (P14) {}; &
    \node[crossx] (P15) {}; &
    \node[phase] (P16) {};  &
        \node[crossx] (P17) {}; &
    \node[phase] (P18) {}; &
    \node[phase] (P19) {};  &
        \node[phase] (P110) {};  &
    \node (P111) {$|a\rangle$}; &[0.5cm] 
    \coordinate (end1); \\

    \node (q2) {$|b\rangle$}; &[-0.5cm] 
    \node[crossx] (P21) {}; &
    \node[] (P22) {}; &
    \node[crossx] (P23) {}; &
        \node[] (P24) {}; &
    \node[crossx] (P25) {}; &
    \node[] (P26) {}; &
        \node[crossx] (P27) {}; &
    \node[] (P28) {}; &
    \node[circlewc] (P29) {}; &
        \node[circlewc] (P210) {}; &
     \node (P211) {$|b\rangle$}; &[0.5cm] 
    \coordinate (end2);\\
    
   \node (q3) {$|c\rangle$}; &
    \node[] (P31) {}; &
    \node[circlewc] (P32) {}; &
   \node[] (P33) {}; &
    \node[circlewc] (P34) {}; &
        \node[] (P35) {}; &
    \node[circlewc] (P36) {}; &
   \node[] (P37) {}; &
    \node[circlewc] (P38) {}; &
       \node[] (P39) {}; &
        \node[] (P310) {}; &
          \node (P311) {$|c\oplus ab\rangle$}; &[-0.5cm] 
    \coordinate (end3); \\
   };

   \draw[thick] (q1) -- (P111)  (q2) -- (P211)  (q3) -- (P311) (P12) -- (P32)  (P11) -- (P21)  (P13) -- (P23) (P14) -- (P34)  (P15) -- (P25) (P16) -- (P36)  (P17) -- (P27)  (P18) -- (P38)  (P19) -- (P29)  (P110) -- (P210);

    \begin{pgfonlayer}{background}
     
       \node[surround] (background) [fit = (q1) (P31) (P311)] {};
               
   \end{pgfonlayer}
    \end{tikzpicture}

\caption{Schematic picture of the decomposition of the Toffoli gate omitting one-qubit gates.  A $SWAP$ gate can be implemented using three $CNOT$ gates, see Fig.\ \ref{figswap}. Note that, in a Toffoli gate, $|a\rangle$ acts as a control qubit for a $CNOT$ operation between $|b\rangle$ and $|c\rangle$, where $|b\rangle$ acts as the control qubit. $|c\oplus ab\rangle$ denotes the sum modulo 2 of $c$ and the product of $a$ and $b$. Further, note the unusual design of the circuit that is such that there is no direct gate (interaction) between $|b\rangle$ and $|c\rangle$. This is due to the physical restrictions in the construction of the elementary gates we are using in this work.} \label{figtoffoli}

\end{subfigure}

\caption{Schematic picture of the decomposition of the $SWAP$ and the Toffoli gate in terms of more elementary gates. } \label{figgates2}
\end{figure}
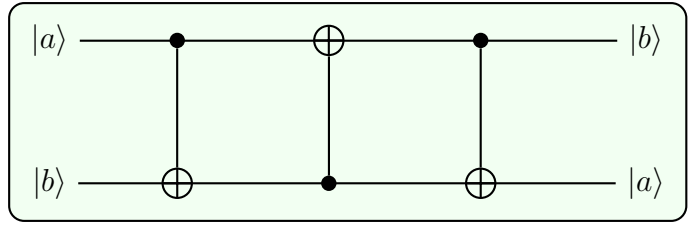
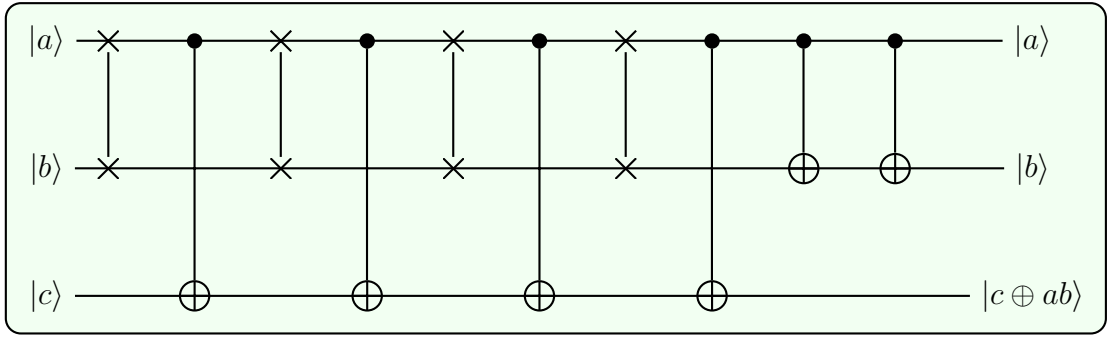

We briefly describe the action of the quantum gates introduced above. More details can be found, for example, in \cite{nielsen2010quantum}. As a basis, we choose the so-called computational basis $\{|0\rangle,\,|1\rangle\}$ that spans the one qubit Hilbert space. Above, we made use of two one-qubit gates. Namely, the single qubit phase gate $R(\frac{\pi}{4})$ that acts, as 
\begin{equation}
|0\rangle\to|0\rangle,\hspace{15mm} |1\rangle\to e^{i \frac{\pi}{4}}|1\rangle
\end{equation}
and the Hadamard gate $H$, whose action is given by 
\begin{equation}
|0\rangle\to\frac{1}{\sqrt{2}}(|0\rangle+|1\rangle),\hspace{15mm} |1\rangle\to \frac{1}{\sqrt{2}}(|0\rangle-|1\rangle).
\end{equation}
Besides single qubit gates there are quantum gates acting on more than one qubit. One  common two qubit gate is the controlled NOT gate, $CNOT$, that takes one control qubit and one qubit as input. If the control qubit is in state $|0\rangle$ the target qubit $|a\rangle$ remains unchanged (the identity operator is applied), while for $|1\rangle$ the target qubit gets flipped
\begin{equation}
|0, a\rangle\to|0, a\rangle,\hspace{15mm} |1, a\rangle\to |1, a\oplus 1\rangle.
\end{equation}
Another useful two qubit gate can be constructed solely from $CNOT$ gates. It is the two qubit gate $SWAP$ that takes two qubits and swaps them as follows
\begin{equation}
|\alpha\rangle\otimes |\beta\rangle\to|\beta\rangle\otimes |\alpha\rangle.
\end{equation}
As mentioned the $SWAP$ gate can be decomposed in terms of (three) $CNOT$ gates, see figure \ref{figswap}. Last, we introduce the controlled $CNOT$ gate that is referred to as Toffoli gate. It is a three qubit gate and, as the $SWAP$ gate, can be constructed from the gates we have discussed so far, see figure \ref{figtoffoli}. The action of this gate on three qubits is trivial whenever the control qubits are not in  state $|1\rangle\otimes |1\rangle$. In the case both control qubits are in state $|1\rangle$ the input qubit gets flipped.

\subsection{Construction of gate Hamiltonians}\label{constrhamiltonians}

The following construction of the gate Hamiltonians for critical qubits coupled to external modes follows the usual construction rules of logic gates. The peculiarity in case of the black hole based quantum critical systems is  that diagonal and non-diagonal terms in the gate Hamiltonian are of the same order.  That is, the gap of the critical qubit, the mixing between the modes and gap of the external mode are all of the same order and are set by the criticality gap $\epsilon$.   The smallness of the mixing and the gap of the external mode are due to the fact that the external mode coupling must preserve criticality. As  explained  in \cite{Dvali:2015wca, Dvali:2016wca} in case of gravity this is  automatically satisfied because the strength of the coupling is gravitational and the frequency of the external mode is red-shifted up to $1/N$ effect.   

 After having constructed the phase gate in Sec.\ \ref{hamforgates}, there is only one one-qubit gate missing (the Hadamard gate $H$) to obtain a universal set of one-qubit gates on the $b-$modes. We find that the (controlled) Hadamard gate is generated by
\begin{equation}\label{ansatzhada}
H_H=c^\dagger c \left(\beta_3+\beta_1  b^\dagger b+\beta_2(b+b^\dagger)\right).
\end{equation}
This can be seen as follows. $H_H$ acts on the basis states as
\begin{align}\label{hamhada}
H_H|0_c 0_b\rangle=& 0, \hspace{20mm}  H_H|1_c 0_b\rangle= \beta_3|1_c 0_b\rangle+\beta_2|1_c 1_b\rangle,\nonumber\\
H_H|0_c 1_b\rangle=& 0,  \hspace{20mm}  H_H |1_c 1_b\rangle=  (\beta_3+\beta_1)|1_c 1_b\rangle+\beta_2|1_c 0_b\rangle .
\end{align}
To compute the time-evolution with respect to Hamiltonian (\ref{ansatzhada}), we restrict ourselves to the non-trivial sector and introduce the notation $a|1_c 0_b\rangle+b|1_c 1_b\rangle=\left(
\begin{array}{c}
a\\
b\\
\end{array}
\right)$. This allows us to write
\begin{equation}\label{eigenhada}
(H_b+H_H) \left(
\begin{array}{c}
a\\
b\\
\end{array}
\right)
=E_H\left(
\begin{array}{c}
a\\
b\\
\end{array}
\right),
\end{equation}
where $E_H$ is given by 
\begin{equation}
E_H= \left(
  \begin{array}{ c c }
     \beta_3 & \beta_2 \\
     \beta_2 & \beta_3+\beta_1+\epsilon
  \end{array} \right),
\end{equation}
where we included the action of the free Hamiltonian $H_b$. Equation (\ref{eigenhada}) allows us to study the time evolution by exponentiating the matrix $E_H$. Thus, the time evolution is given by
\begin{equation}\label{timehada}
\left(
\begin{array}{c}
a\\
b\\
\end{array}
\right)(t)=e^{-i E_H t}\left(
\begin{array}{c}
a\\
b\\
\end{array}
\right).
\end{equation}
Calculating (\ref{timehada}), we find the following expression for the time evolved state
\begin{equation}\label{timehada2}
e^{-it(\beta_3+\frac{\beta_1+\epsilon}{2})}\left(
\begin{array}{c}
a \cos(\sqrt{\beta_2^2+\frac{(\beta_1+\epsilon)^2}{4}} t)+i \sin(\sqrt{\beta_2^2+\frac{(\beta_1+\epsilon)^2}{4}} t) \frac{a\frac{\beta_1+\epsilon}{2}-b \beta_2}{\sqrt{\beta_2^2+\frac{(\beta_1+\epsilon)^2}{4}} }\\
b \cos(\sqrt{\beta_2^2+\frac{(\beta_1+\epsilon)^2}{4}} t)-i \sin(\sqrt{\beta_2^2+\frac{(\beta_1+\epsilon)^2}{4}} t) \frac{b\frac{\beta_1+\epsilon}{2}+a \beta_2}{\sqrt{\beta_2^2+\frac{(\beta_1+\epsilon)^2}{4}} }\\
\end{array}
\right).
\end{equation}
Using that the action of the Hadamard gate is achieved by $H_b+H_H$ if after some time $t$ the input states are transformed as
\begin{equation}
\left(
\begin{array}{c}
1\\
0\\
\end{array}
\right)\longrightarrow \,\frac{1}{\sqrt{2}}\left(
\begin{array}{c}
1\\
1\\
\end{array}
\right),\hspace{10mm}\left(
\begin{array}{c}
0\\
1\\
\end{array}
\right)\longrightarrow\,
\frac{1}{\sqrt{2}}\left(
\begin{array}{c}
1\\
-1\\
\end{array}
\right),
\end{equation}
we obtain a system of four equations for the four variables $\beta_1, \beta_2, \beta_3$ and the time $t$. However, only two of them are independent. These are,
\begin{equation}\label{equations}
(-1)^n= i   \frac{(\beta_1+\epsilon)e^{-it(\beta_3+\frac{\beta_1+\epsilon}{2})} }{\sqrt{2}\sqrt{\beta_2^2+\frac{(\beta_1+\epsilon)^2}{4}} },\hspace{10mm}
(-1)^{n+1}= i \sqrt{2}    \frac{\beta_2 e^{-it(\beta_3+\frac{\beta_1+\epsilon}{2})}}{\sqrt{\beta_2^2+\frac{(\beta_1+\epsilon)^2}{4}} },
\end{equation}
where we already set $t\sqrt{\beta_2^2+\frac{(\beta_1+\epsilon)^2}{4}} =(n+\frac{1}{2})\pi$, where $n\in \mathbb{Z}$. Furthermore, we fix $\beta_3$ such that $t(\beta_3+\frac{\beta_1+\epsilon}{2})=\frac{3\pi}{2}$. Then, for $n=0$ equations (\ref{equations}) reduce to
\begin{equation}\label{equations2}
\sqrt{2} = -   \frac{\beta_1+\epsilon}{\sqrt{\beta_2^2+\frac{(\beta_1+\epsilon)^2}{4}} },\hspace{10mm}
\frac{1}{\sqrt{2} }=   \frac{\beta_2}{\sqrt{\beta_2^2+\frac{(\beta_1+\epsilon)^2}{4}} },
\end{equation}
that are satisfied for $\beta_2=-\frac{\beta_1+\epsilon}{2}$. There is one further constraint that we ignored so far. From (\ref{hamhada}) it is evident that $H_H$ only affects states that have non-vanishing occupation of the $c-$mode. However, the free evolution with respect to $H_b$ introduces a phase $\phi=\epsilon t$ for $|0_c 1_b\rangle$ that naturally constrains $t$ according to $t=\frac{2\pi l}{\epsilon}$, where $l\in \mathbb{Z}$. For $l=1$, we find
\begin{equation}
\beta_1=-\epsilon (1+\frac{1}{2\sqrt{2}}),   \hspace{10mm} \beta_2=\frac{\epsilon}{4\sqrt{2}},   \hspace{10mm} \beta_3=\frac{\epsilon}{4}(3+\frac{1}{\sqrt{2}}).
\end{equation}
 Next, we discuss the $CNOT$ gate controlled by external $c-$modes that was found in \cite{Dvali:2015wca}. The generating Hamiltonian is given by
\begin{equation}\label{ansatzcnotdv}
H_{CNOT_{cb}}=\nu b^\dagger b (c^\dagger +c)+\delta c^\dagger c,
\end{equation}
where $\nu\sim\epsilon$, $\delta\to 0$ and the time $t$ required to implement the gate is $t=\frac{\pi}{2\nu}\sim \frac{\pi}{2\epsilon}$. The last gate we have to implement is  the $CNOT$ gate with the roles of $b-$ and $c-$modes reversed. Using the same ansatz (\ref{ansatzhada}) as for the Hadamard gate
\begin{equation}\label{ansatzcnot}
H_{CNOT_{bc}}=c^\dagger c \left(\gamma_3+\gamma_1  b^\dagger b+\gamma_2(b+b^\dagger)\right).
\end{equation}
where for clarity we changed the notation to $\gamma_i$ instead of $\beta_i$. With $\left(
\begin{array}{c}
a\\
b\\
\end{array}
\right)$ defined as above, we can write
\begin{equation}\label{eigencnot}
(H_b+H_{CNOT_{bc}}) \left(
\begin{array}{c}
a\\
b\\
\end{array}
\right)
=\left(
  \begin{array}{ c c }
     \gamma_3 & \gamma_2 \\
     \gamma_2 & \gamma_3+\gamma_1+\epsilon
  \end{array} \right)\left(
\begin{array}{c}
a\\
b\\
\end{array}
\right).
\end{equation}
The expression for the time evolved state coincides with (\ref{timehada2}) after the substitution $\beta_i\to\gamma_i$. However, now we request the following final states after a time evolution for a time $t$
\begin{equation}
\left(
\begin{array}{c}
1\\
0\\
\end{array}
\right)\longrightarrow \,\left(
\begin{array}{c}
0\\
1\\
\end{array}
\right),\hspace{10mm}
\left(
\begin{array}{c}
0\\
1\\
\end{array}
\right)\longrightarrow\,
\left(
\begin{array}{c}
1\\
0\\
\end{array}
\right).
\end{equation}
Again, we obtain two linearly independent equations. These are given by
\begin{equation}\label{equationscnot}
0=  \frac{\gamma_1+\epsilon}{\sqrt{\gamma_2^2+\frac{(\gamma_1+\epsilon)^2}{4}} },\hspace{10mm}
1= -i    \frac{\gamma_2 e^{-it(\gamma_3+\frac{\gamma_1+\epsilon}{2})} }{\sqrt{\gamma_2^2+\frac{(\gamma_1+\epsilon)^2}{4}} },
\end{equation}
where we already set $t\sqrt{\gamma_2^2+\frac{(\gamma_1+\epsilon)^2}{4}} =\frac{\pi}{2}$. It is evident that $\gamma_1=-\epsilon$ and $\gamma_3 t=\frac{3\pi}{2}$. The constraint coming from the triviality of the time evolution of $|0_c 1_b\rangle$ gives rise to $t=\frac{2\pi l}{\epsilon}$, where $l\in \mathbb{Z}$. Thus, there are three conditions on $t$: $t=\frac{2\pi l}{\epsilon}$, $ t=\frac{3\pi }{2\gamma_3}$, $ t =\frac{\pi}{\gamma_2}$ and we find
\begin{align}
\gamma_1=-\epsilon, \hspace{10mm}  \gamma_2=\frac{\epsilon}{2}, \hspace{10mm}  \gamma_3=\frac{3}{4}\epsilon,
\end{align}
for $l=1$. Finally, we need to implement an entangling two-qubit gate between two Bogoliubov modes $b_i$ and $b_j$. We use a $c-$mode as the ``mediator'' of the interaction, just as before. More precisely, we implement a Toffoli gate. This can be done using the gates we already obtained above, see Fig.\ \ref{figtoffoli}. Alternatively, we can use an (additional) external mode as memory to perform the gate. As only a constant number of gates is required to construct a circuit that implements the Toffoli gate, the time required to apply it is of the same order than for the gates we discussed so far. This ends the construction of a universal set of gates.

\bibliographystyle{JHEP}
\tocless\bibliography{literatureBH}

\end{document}